\documentclass[aps,prc,twocolumn,superscriptaddress]{revtex4}

\usepackage{subfig}
\usepackage{graphicx}
\begin{document}


\title{A new inverse quasifission mechanism to produce neutron-rich transfermium nuclei}


\author{David J. Kedziora}
\affiliation{Department of Nuclear Physics, Research School of Physics and Engineering, Australian National University, Canberra, Australian Capital Territory 0200, Australia}

\author{C\'{e}dric Simenel}\email[]{cedric.simenel@cea.fr}
\affiliation{Department of Nuclear Physics, Research School of Physics and Engineering, Australian National University, Canberra, Australian Capital Territory 0200, Australia}\affiliation{CEA, Centre de Saclay, IRFU/Service de Physique Nucl\'eaire, F-91191 Gif-sur-Yvette, France.}

\date{\today}

\begin{abstract}
Based on time-dependent Hartree-Fock theory, a new inverse quasifission mechanism is proposed to produce neutron-rich transfermium nuclei, in collision of prolate deformed actinides. Calculations show that collision of the tip of one nucleus with the side of the other results in a nucleon flux toward the latter. The role of nucleon evaporation and impact parameter, as well as the collision time are discussed.
\end{abstract}

\pacs{}

\maketitle

\section{Introduction}
\label{sec:intro}

The quest for, and study of the heaviest elements has involved much experimental and theoretical effort in recent years.
Their existence relies only on stabilizing quantum shell effects,
which make them ideal to test quantum mechanical nuclear structure models. 
Both the location of the predicted island of  stability in the superheavy element (SHE) region~\cite{ben99,ben01,mor08}
and spectroscopy of transfermium nuclei ($Z>100$) \cite{cha06,her06} are needed to constrain these models.
A natural way to form such nuclei is through fusion of heavy nuclei,
followed by neutron and gamma evaporation from the compound nucleus. 
SHEs have  been produced either in ''cold'' fusion reactions based on closed shell target 
nuclei~\cite{hof00,mor07}, or in ''hot'' fusion reactions involving actinide targets~\cite{oga06,hof07}.
The heaviest element, containing 118 protons, has been synthesized with the latter technique~\cite{oga06}. 
However, $\alpha$-decay chains of SHEs formed in hot fusion end in a region of unknown neutron-rich isotopes
with 104-110 protons. 
Thus, it is necessary to study this region of the nuclear chart in order to provide 
a better identification of the decay daughters and confirm these SHEs.

Fusion-evaporation cross-sections decrease rapidly with the product of the charges of the reactants,
down to few picobarns for SHEs. 
These cross sections are too small to allow a detailed study of nuclear structure. 
For instance, a basic property like the mass has been measured only recently 
for $^{252-254}$No fusion products with a Penning trap mass spectrometer~\cite{blo10}.
Furthermore, fusion reactions lead usually to neutron deficient compound systems, 
which, in addition, decay by neutron emission. 
It is therefore worth exploring other reaction mechanisms to produce and study the heaviest nuclei,
and, in particular, their neutron-rich isotopes.

An alternative way to form neutron-rich heavy nuclei is to consider multinucleon transfer 
in such a way that one ejectile gets heavier that any of the reactants~\cite{vol77}.
This process is sometimes called "asymmetry-exit-channel"~\cite{ada05} 
or "inverse"~\cite{zag06} quasifission, as the mass asymmetry of the outgoing fragments
has increased, whereas "standard" quasifission tends to reduce this asymmetry. 
Such a process has been investigated experimentally considering 
either a light-medium mass projectile on an actinide target~\cite{lee82,lee83,gag86,hof85,tur92},
or actinide collisions~\cite{sch78,fre79,sch82,kra86,gol10}.
Recent theoretical studies of multinucleon transfer have been performed in the 
dinuclear system model (DNS)~\cite{ada05,fen09}, 
using multidimensional Langevin equations~\cite{zag06,zag07,zag08a,zag08b},
in the constrained molecular dynamics model~\cite{mar02},
in the improved quantum molecular dynamics approach~\cite{tia08,zha09},
and within the time-dependent Hartree-Fock (TDHF) theory~\cite{gol09}.
In particular, it is predicted that shell effects in the $^{208}$Pb region should favor inverse quasifission~\cite{fre79,zag06,fen09}.
Indeed,  as one actinide falls into the valley of the potential energy surface toward
the magic numbers $Z=82$ and $N=126$, the mass and charge of its collision partner increases correspondingly.

Multinucleon transfer depends also strongly on deformation 
and relative orientation of the nuclei~\cite{gol09,zha09}.
This should play an important role in actinide collisions as nuclei have strong prolate deformations 
in this region of the nuclear chart~\cite{hil07}.
In particular, it has been shown, in the case of the symmetric central collision $^{238}$U+$^{238}$U, 
that a nucleon flux appears in the neck when one nucleus has its deformation axis aligned with the collision axis
and perpendicular to the deformation axis of the collision partner~\cite{gol09}.  
In this case, the "aligned" nucleus loses nucleons. 
The main goal of the present paper is to investigate a new inverse quasifission mechanism 
due  to such orientation effect in initially mass and charge asymmetric collisions of actinides.
As an illustration, we perform calculations for the $^{232}$Th+$^{250}$Cf reaction within the TDHF framework.
In section~\ref{sec:theory}, we present briefly the TDHF theory and give some numerical details of the calculations.
Then, the results are presented and discussed in section~\ref{sec:results}.
Finally, we conclude in section~\ref{sec:conclusion}.

\section{The time-dependent Hatree-Fock approach}
\label{sec:theory}

\subsection{Theory}

The TDHF theory has been proposed by Dirac~\cite{dir30} and 
applied in nuclear physics~\cite{eng75,bon76}, including actinide collisions~\cite{str83},
with Skyrme effective interactions~\cite{sky56}.
In its Liouville form, the TDHF equation is written 
\begin{equation}
\label{eq:TDHF_Liouville}
    i \hbar \frac{\partial \rho}{\partial t} = [h[\rho],\rho].
\end{equation}
It gives the evolution of the one-body density matrix~$\rho$ 
assuming that the system is always described by an antisymmetrized independent particle wave function
to ensure an exact treatment of the Pauli principle during time evolution~\cite{sim10}.
The one-body density matrix can be used to compute expectation values of any one-body observable
and its evolution, within TDHF, accounts for one-body dissipation mechanisms.
The latter are known to drive low-energy reaction mechanisms
as the Pauli blocking prevents nucleon-nucleon collisions.

The one-body density matrix of an independent particle state can be written
\begin{equation}
\label{Eq:Density}
    \rho(\mathbf{r}sq,\mathbf{r'}s'q')=\sum_{i=1}^{A_1+A_2} \varphi_i^*(\mathbf{r'}s'q') \varphi_i(\mathbf{r}sq),
\end{equation}
where $\{\varphi_i\}$ are the occupied single particle wave functions, 
$A_1$ and $A_2$ are the number of nucleons in each nucleus, 
and $\mathbf{r}$, $s$, and $q$ denote the nucleon position, spin and isospin, respectively. 
The single-particle Hartree-Fock Hamiltonian $h[\rho]$ is self-consistent and can be expressed as
\begin{equation}
\label{Eq:Hamiltonian}
    h[\rho](\mathbf{r}sq,\mathbf{r'}s'q')=\frac{\delta E[\rho]}{\delta\rho(\mathbf{r'}s'q',\mathbf{r}sq)},
\end{equation}
where $E[\rho]$ is the Skyrme energy density functional (EDF)
modeling the interaction between the nucleons. 
The EDF is the only phenomenological ingredient, 
as it has been adjusted to reproduce nuclear structure properties~\cite{cha98}.
In practice, the TDHF equation~(\ref{eq:TDHF_Liouville}) is written 
as a set of nonlinear Schr{\"o}dinger-like equations 
for the occupied single-particle wave functions
\begin{equation}
\label{eq:TDHF_sp}
i \hbar \frac{\partial  \varphi_i (t)}{\partial t} = h[\rho (t)] \varphi_i (t).
\end{equation}
Realistic TDHF calculations in 3 dimensions are now possible 
with modern Skyrme functionals including spin-orbit term~\cite{kim97,nak05,uma06,mar06}
and supercomputers allow simulation of realistic actinide collisions~\cite{gol09}.

\subsection{Numerical details}
\label{subsec:numerical}

The nuclei are assumed to be initially in their Hartree-Fock (HF) ground state.
The HF and TDHF calculations are both performed with the SLy4$d$ Skyrme EDF~\cite{kim97},
allowing for a fully consistent treatment of nuclear structure and dynamics. 
HF ground states are generated by solving the stationary version of Eq.~(\ref{eq:TDHF_Liouville}), 
in which the left-hand side is replaced with $0$, 
by using the imaginary-time method~\cite{dav80}. 
The wavefunctions are decomposed in a cartesian basis with a mesh-size unit $\Delta~x=0.8$~fm~\cite{kim97}.
The encapsulating box has to be large enough so that the tails of $\varphi_i$ are not significantly affected 
by the hard-box boundary condition. 
The HF calculations are converged for the $^{250}$Cf nucleus with $16$ steps of $\Delta~x$ from the center of the nucleus.

\begin{figure}[h]
\begin{center}
\includegraphics[width=0.16\textwidth, angle=270]{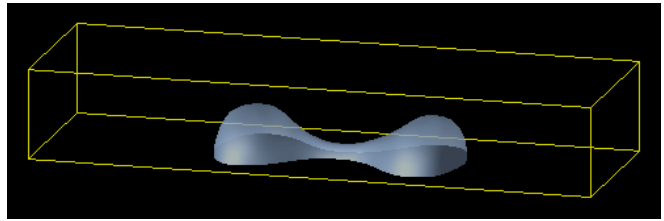}
\end{center}
\caption{\label{Fig:CollisionBox} (color online).  A $12.8$ fm $\times$ $25.6$ fm $\times$ $76.8$ fm half-box 
used for  $^{232}$Th+$^{250}$Cf  central collisions. 
The surface represents an example of isodensity, at half the saturation density $\rho_0/2=0.08$~fm$^{-3}$, of the fragments moving apart after contact.}
\end{figure}

The dynamical calculations for central collisions are performed in a half-box with $N_x=96$, $N_y=32$, and $N_z=16$ mesh-points along the $x$, $y$, and $z$-axis, respectively. The $z=0$ plane is assumed to be a plane of symmetry to speed up the calculations. For non-central collisions, $N_y$ is doubled to allow full re-separation of the fragments without spurious reflections at the box boundaries. The nuclei start initially along the $x$ axis at a distance $D_0=51.2$~fm.
Their initial velocity vectors are determined assuming a Rutherford trajectory 
and they are given a boost by applying a translation in momentum space \cite{tho62}. 
Equations~(\ref{eq:TDHF_sp}) are then solved iteratively 
using a real time propagation algorithm that ensures energy conservation~\cite{bon76,flo78}  
(see also Ref.~\cite{sim10} for more details). 
The {\textsc{tdhf3d}} code~\cite{kim97} is used with time step of $1.5\times10^{-24}$~s 
for a maximum simulation time of $6~\times10^{-21}$~s, 
sufficient for contact and subsequent re-separation of the fragments.
Figure~\ref{Fig:CollisionBox} shows the half-box encapsulating an example of isodensity obtained after  contact of the nuclei
in a central collision.

\section{Multinucleon transfer in $^{232}$Th+$^{250}$Cf  at low-energy}
\label{sec:results}

Let us now study the multinucleon transfer mechanism in the $^{232}$Th+$^{250}$Cf reaction at energies between $626.6$~MeV (no contact) and $1205$~MeV with the {\textsc{tdhf3d}} code.

\subsection{Definition of the relative orientations}

\begin{figure}[h]
\begin{center}
\subfloat{\includegraphics[width=0.05\textwidth, angle=90]{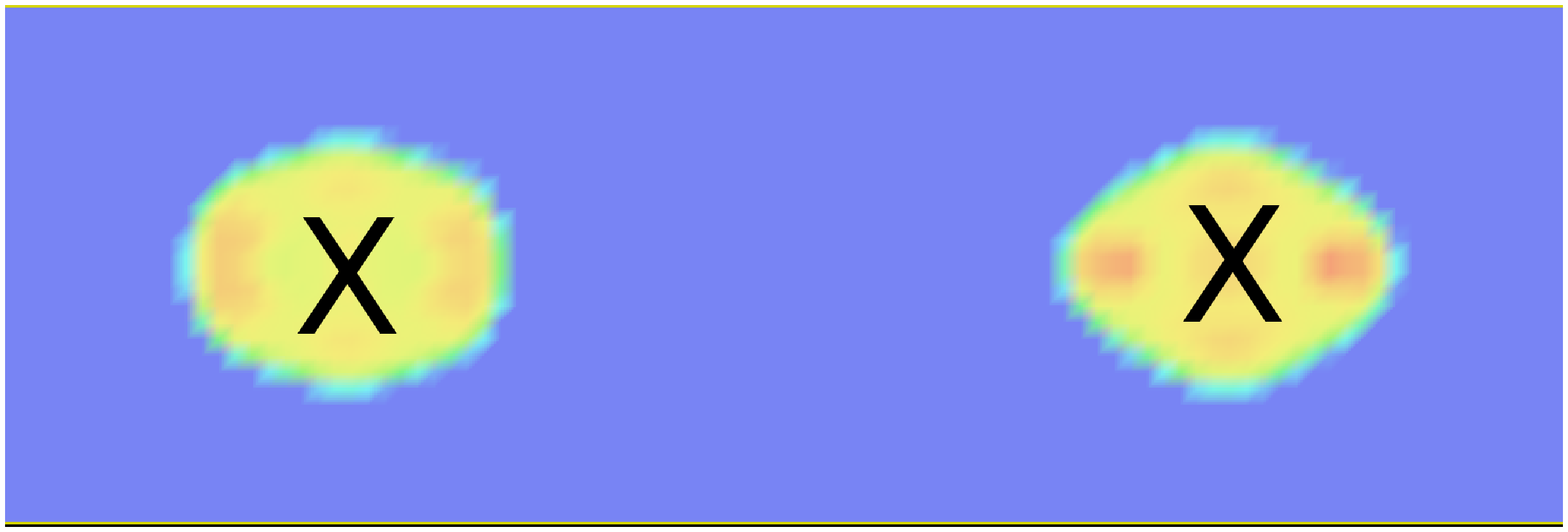}}
\subfloat{\includegraphics[width=0.05\textwidth, angle=90]{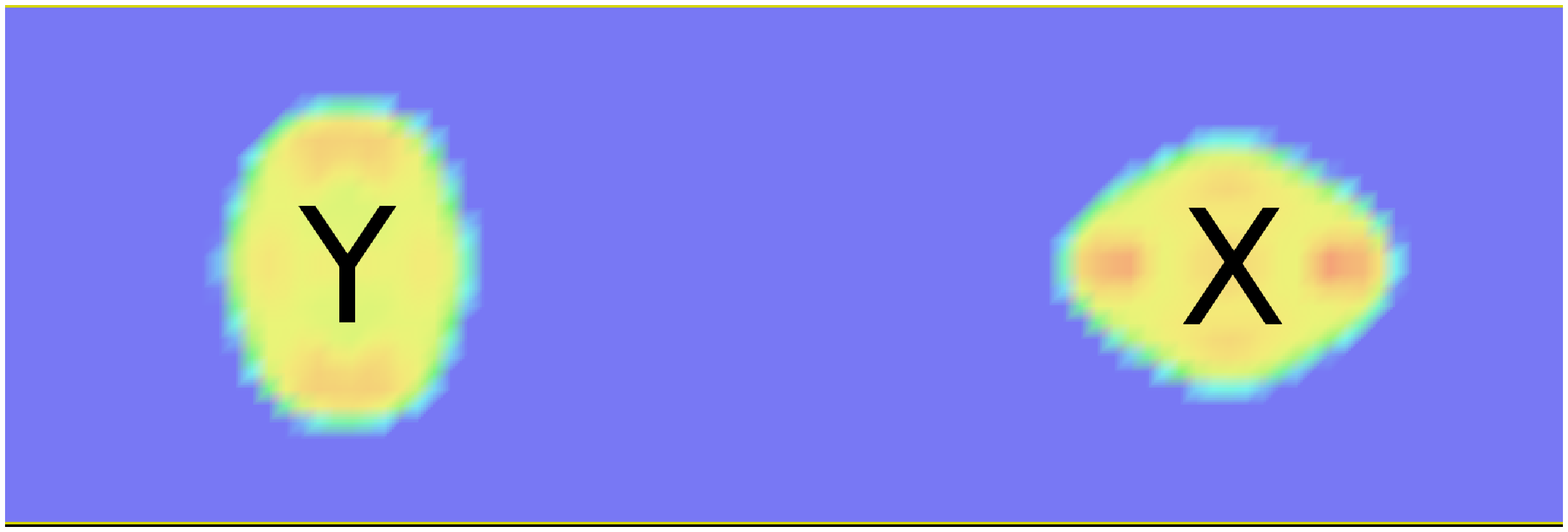}}
\subfloat{\includegraphics[width=0.05\textwidth, angle=90]{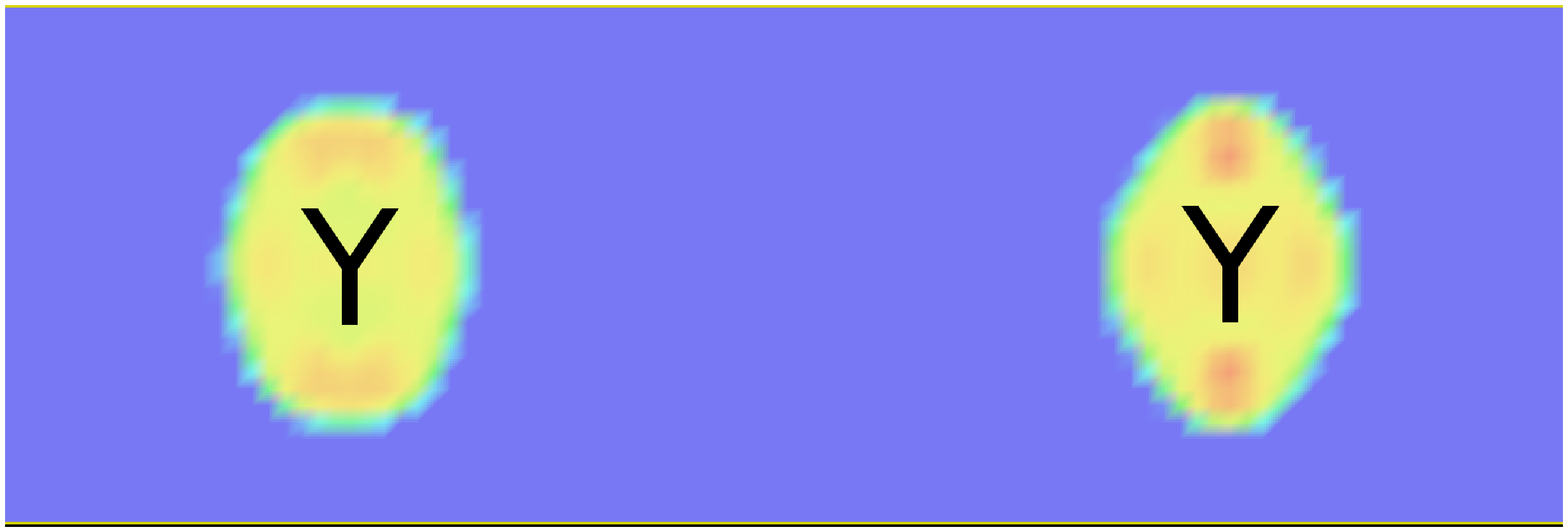}}\\
\subfloat{\includegraphics[width=0.05\textwidth, angle=90]{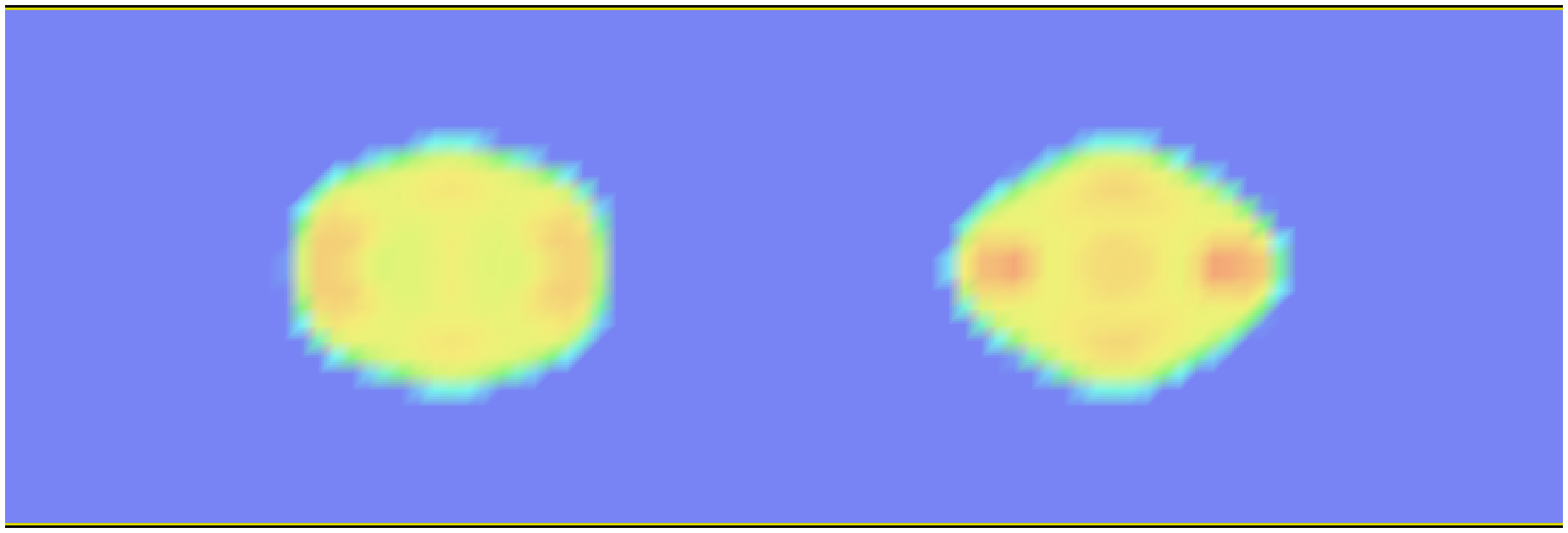}}
\subfloat{\includegraphics[width=0.05\textwidth, angle=90]{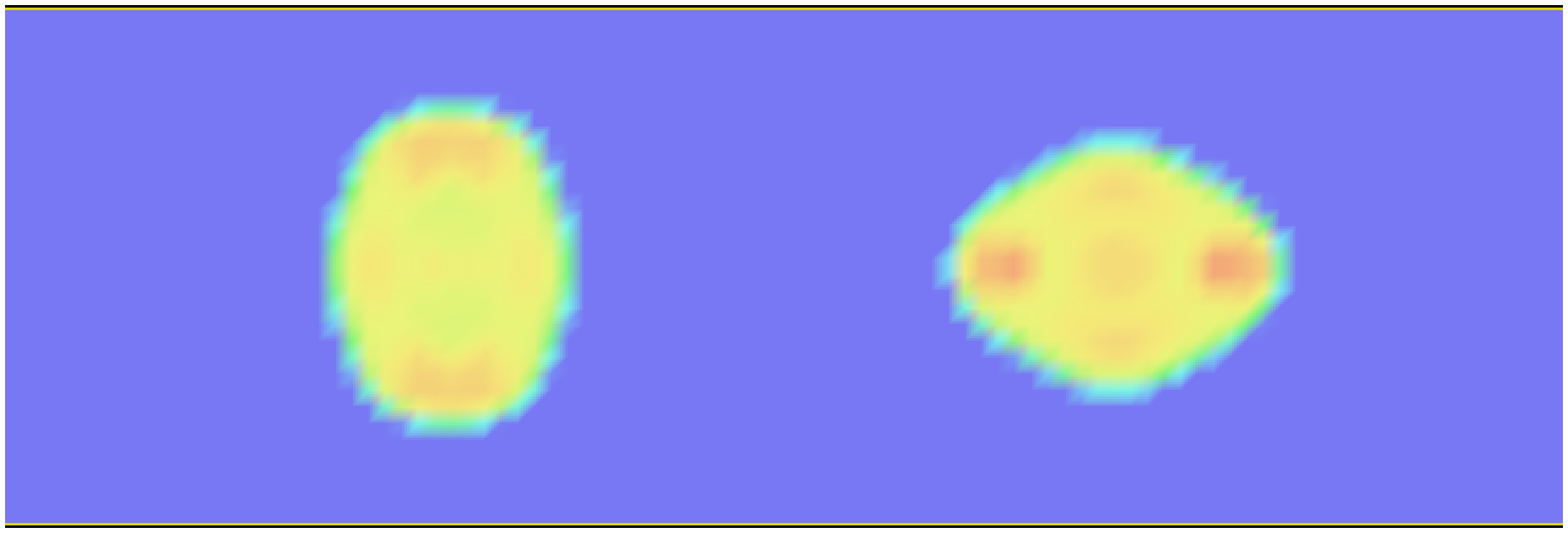}}
\subfloat{\includegraphics[width=0.05\textwidth, angle=90]{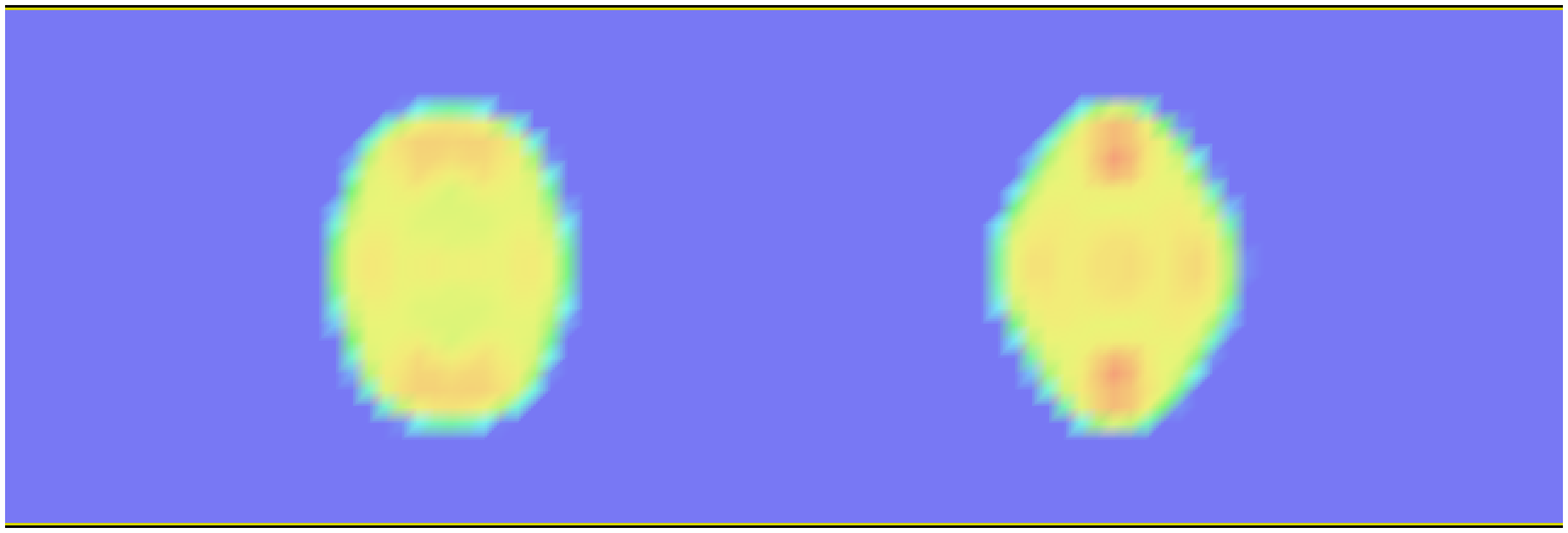}}\\
\subfloat{\includegraphics[width=0.05\textwidth, angle=90]{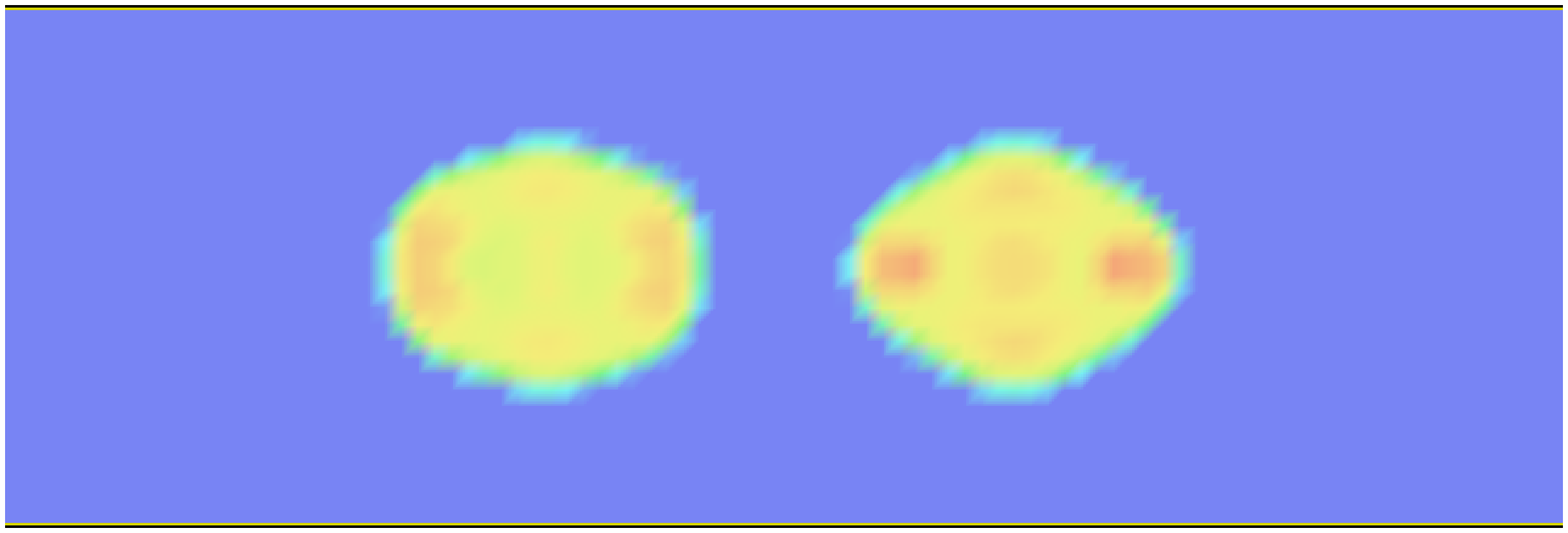}}
\subfloat{\includegraphics[width=0.05\textwidth, angle=90]{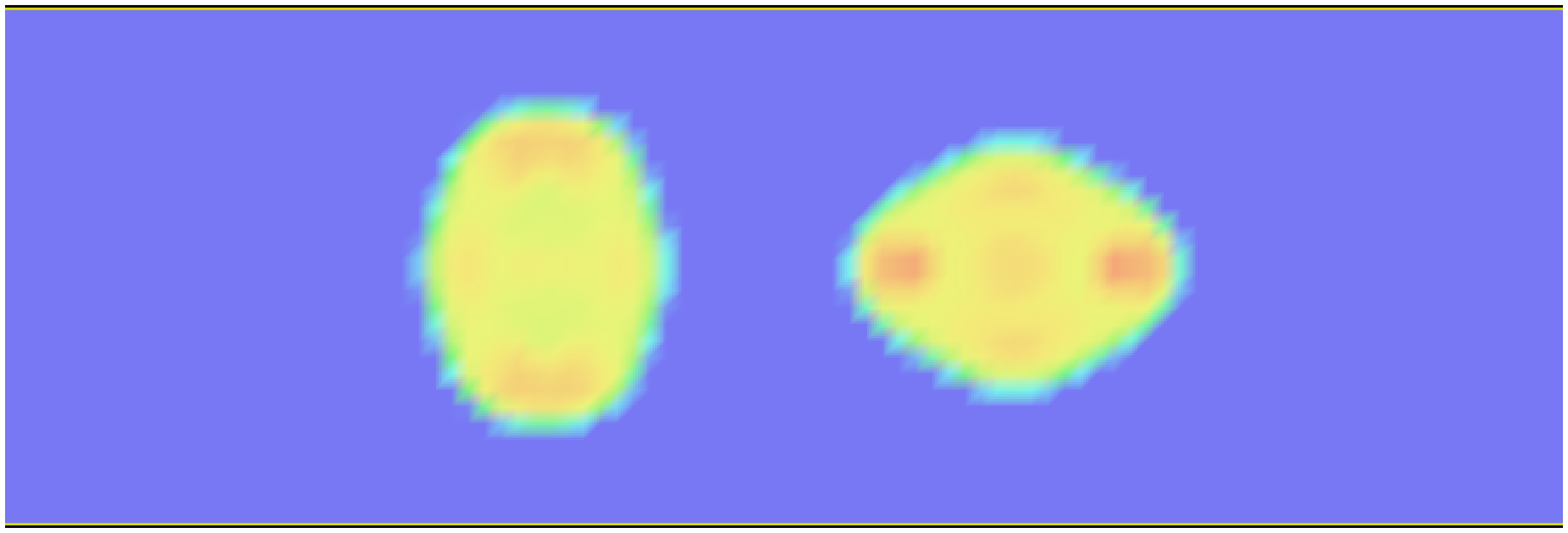}}
\subfloat{\includegraphics[width=0.05\textwidth, angle=90]{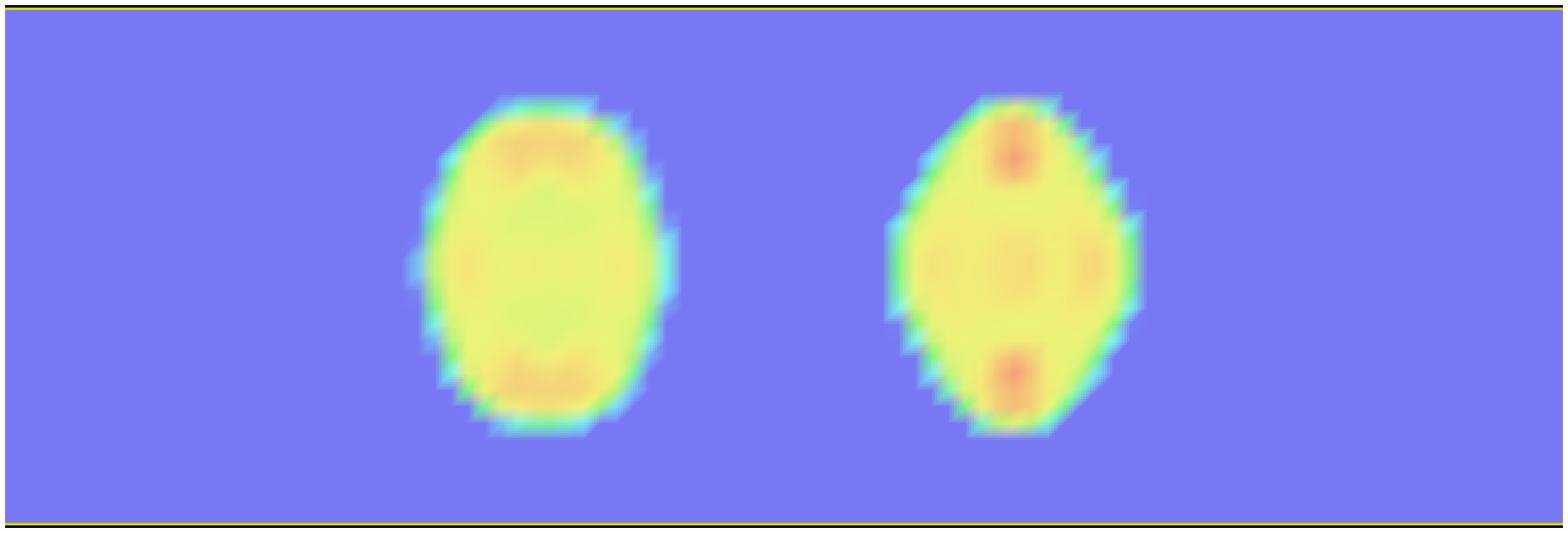}}\\
\subfloat{\includegraphics[width=0.05\textwidth, angle=90]{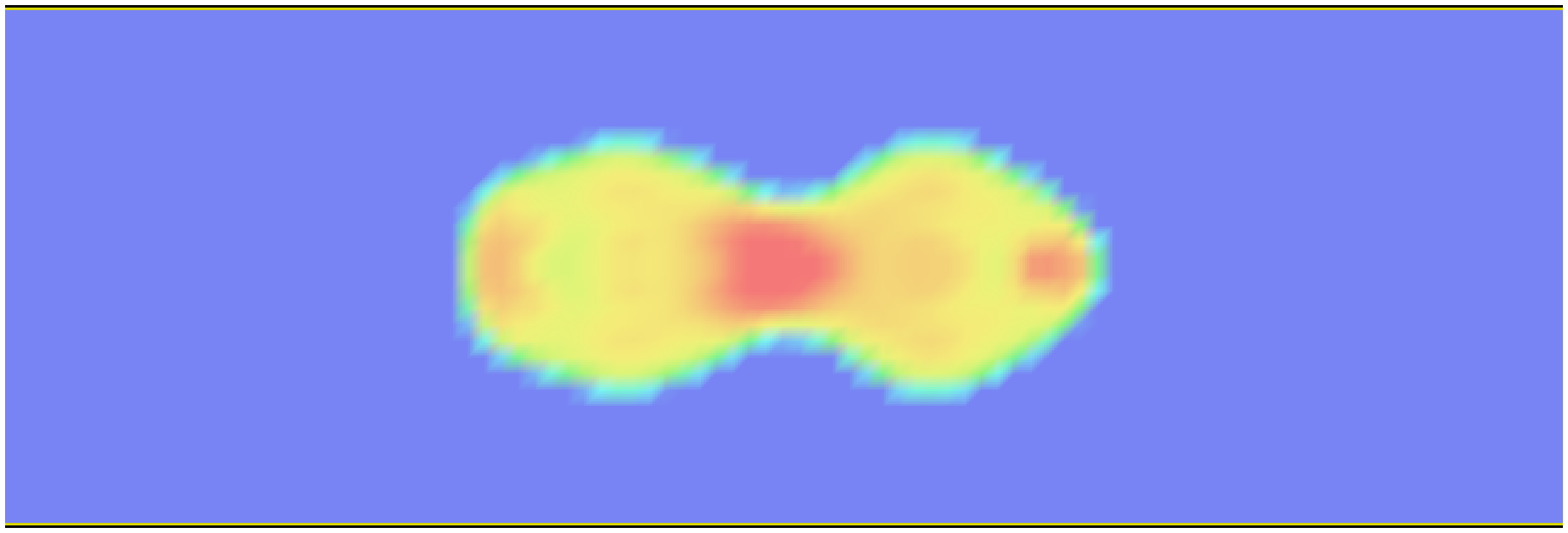}}
\subfloat{\includegraphics[width=0.05\textwidth, angle=90]{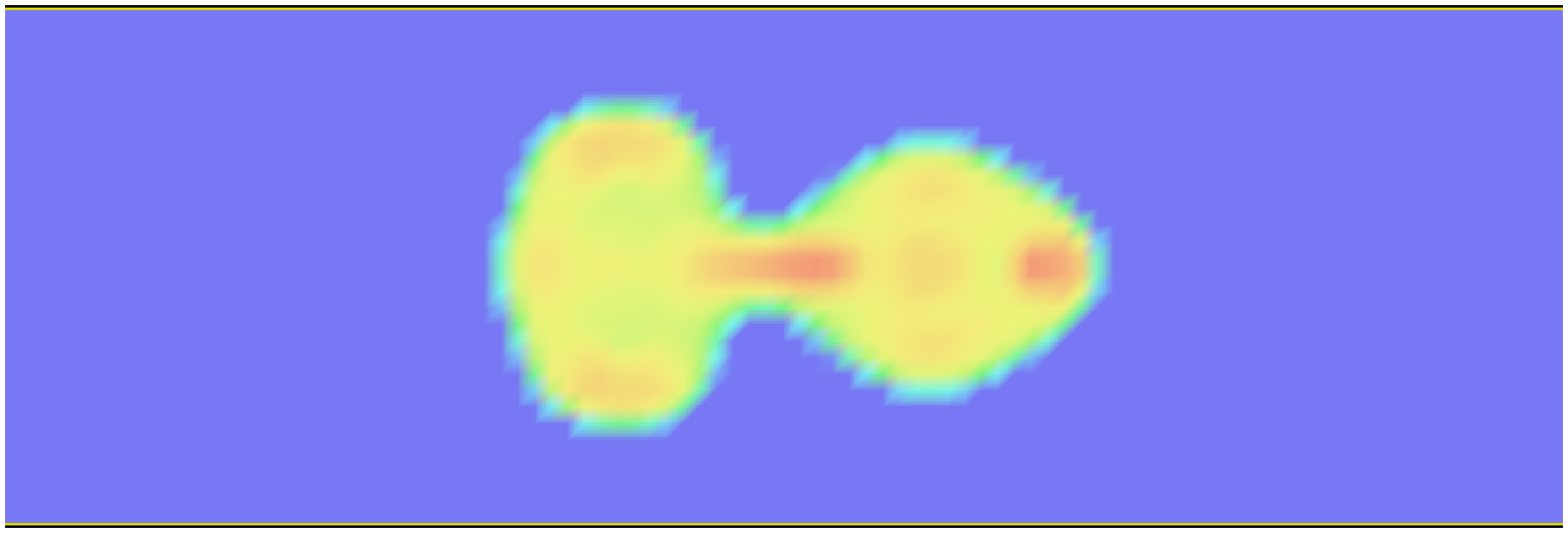}}
\subfloat{\includegraphics[width=0.05\textwidth, angle=90]{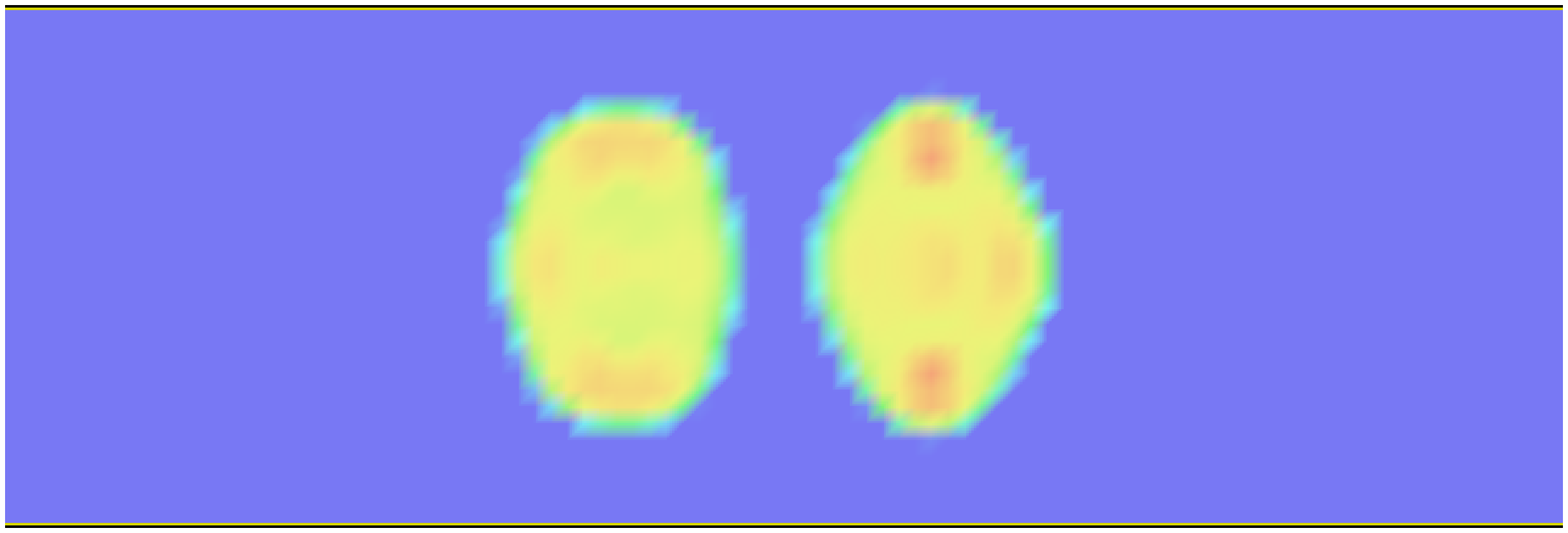}}\\
\subfloat{\includegraphics[width=0.05\textwidth, angle=90]{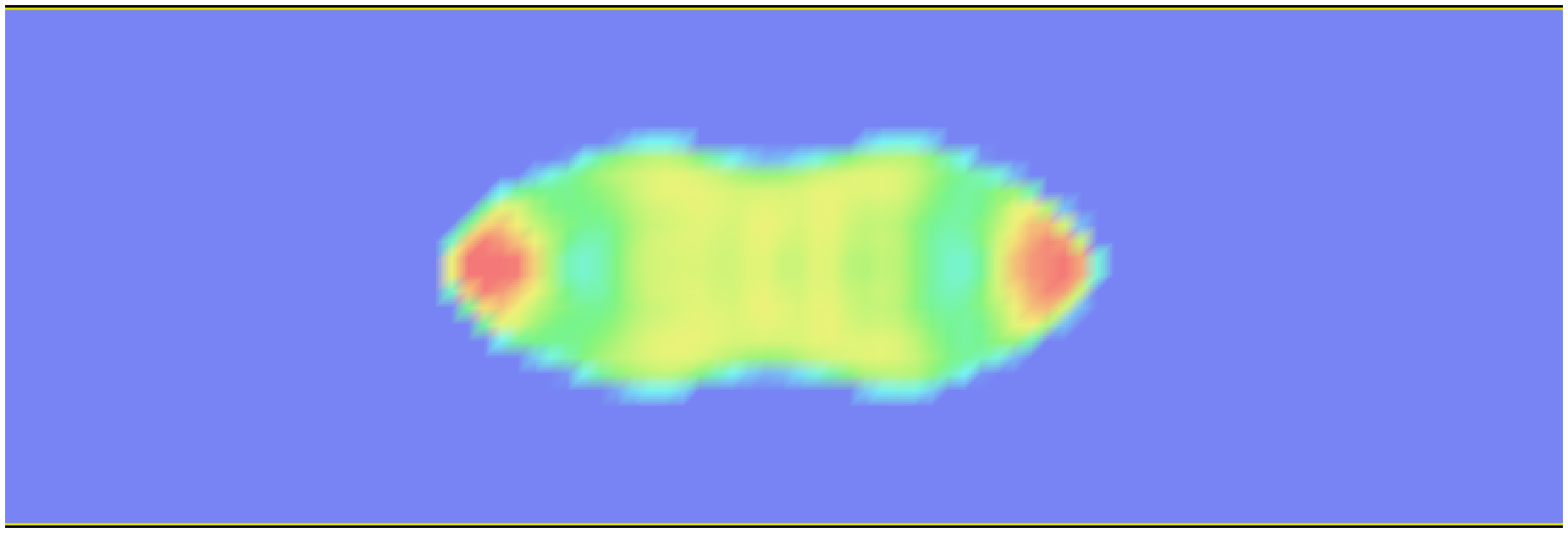}}
\subfloat{\includegraphics[width=0.05\textwidth, angle=90]{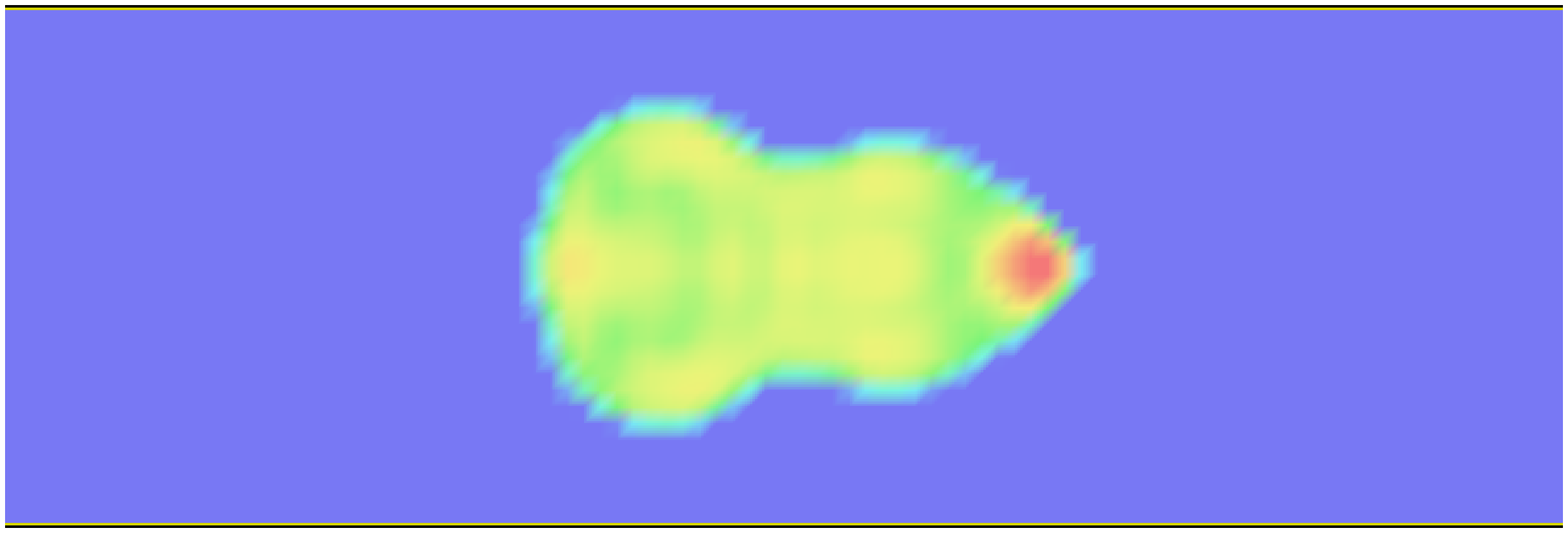}}
\subfloat{\includegraphics[width=0.05\textwidth, angle=90]{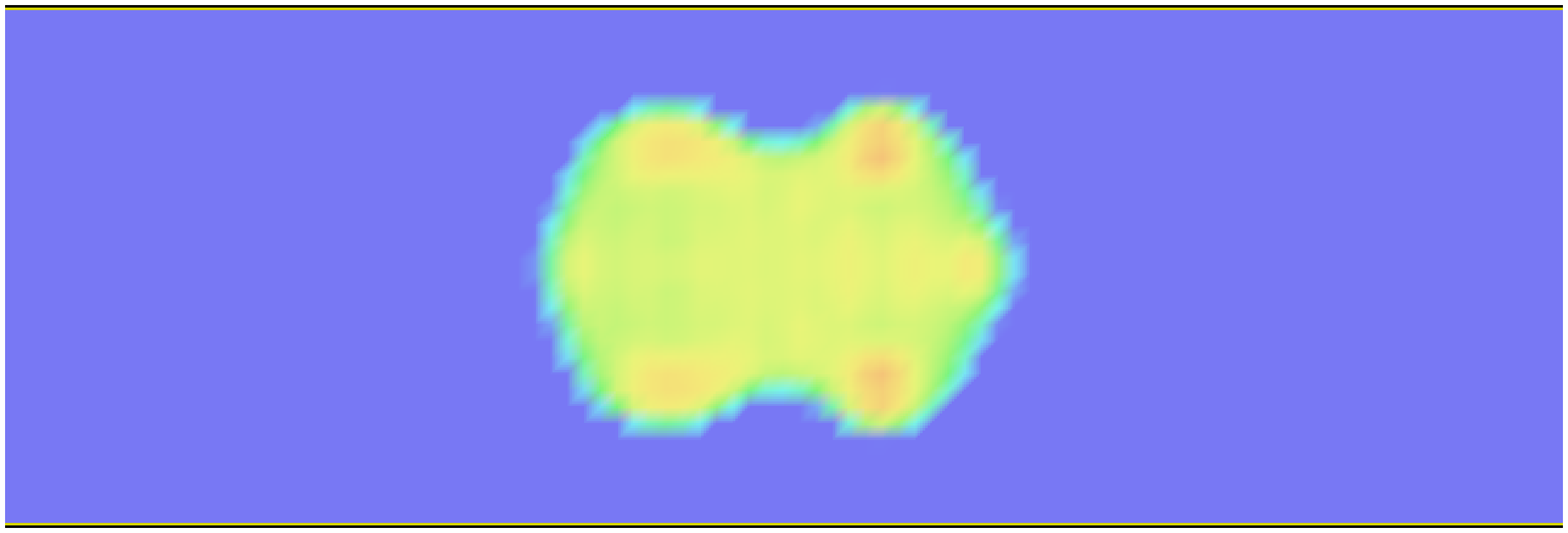}}\\
\subfloat{\includegraphics[width=0.05\textwidth, angle=90]{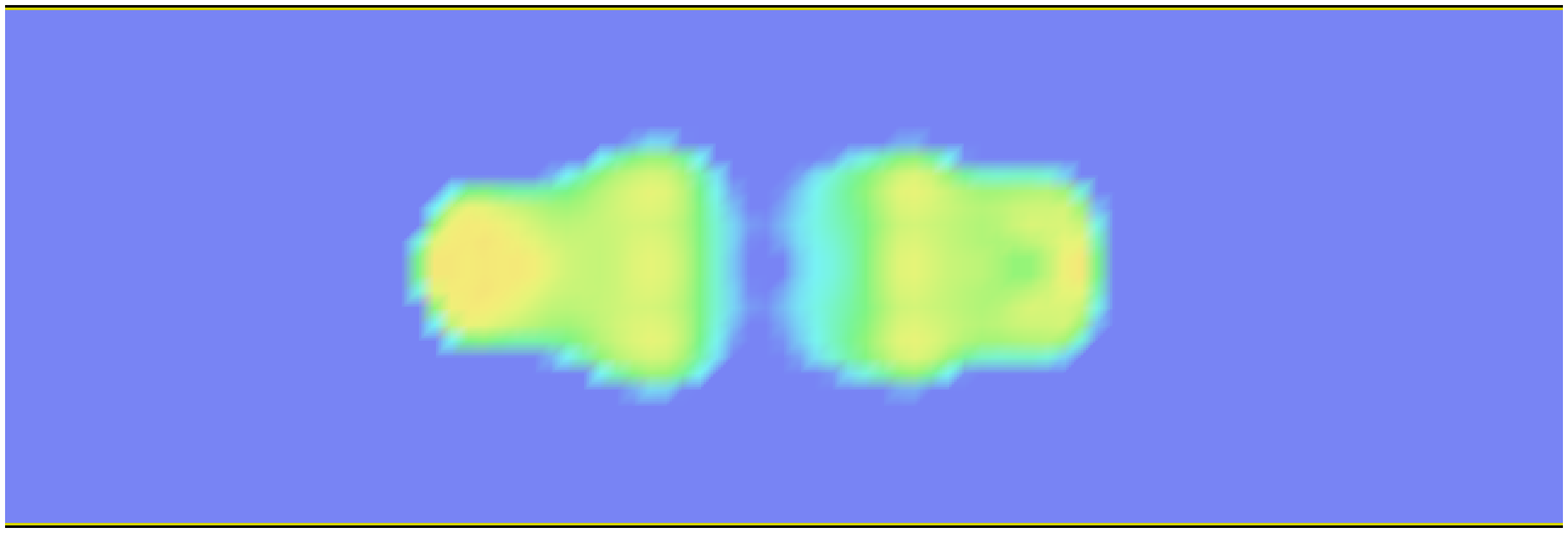}}
\subfloat{\includegraphics[width=0.05\textwidth, angle=90]{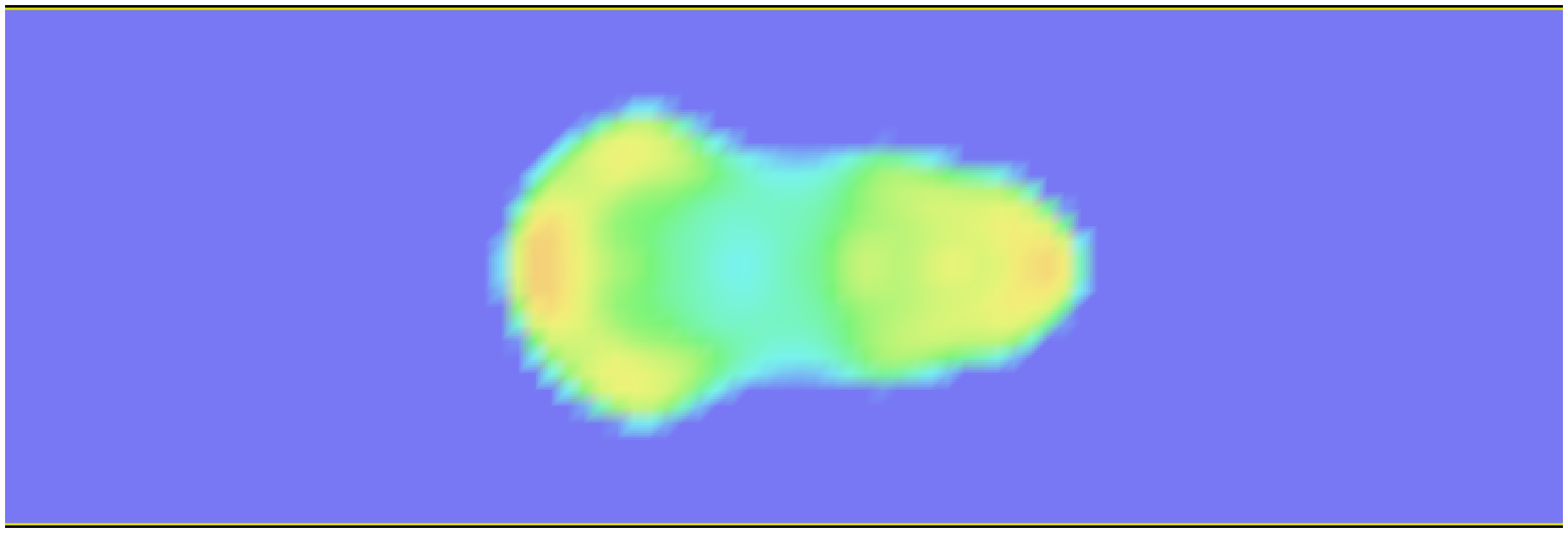}}
\subfloat{\includegraphics[width=0.05\textwidth, angle=90]{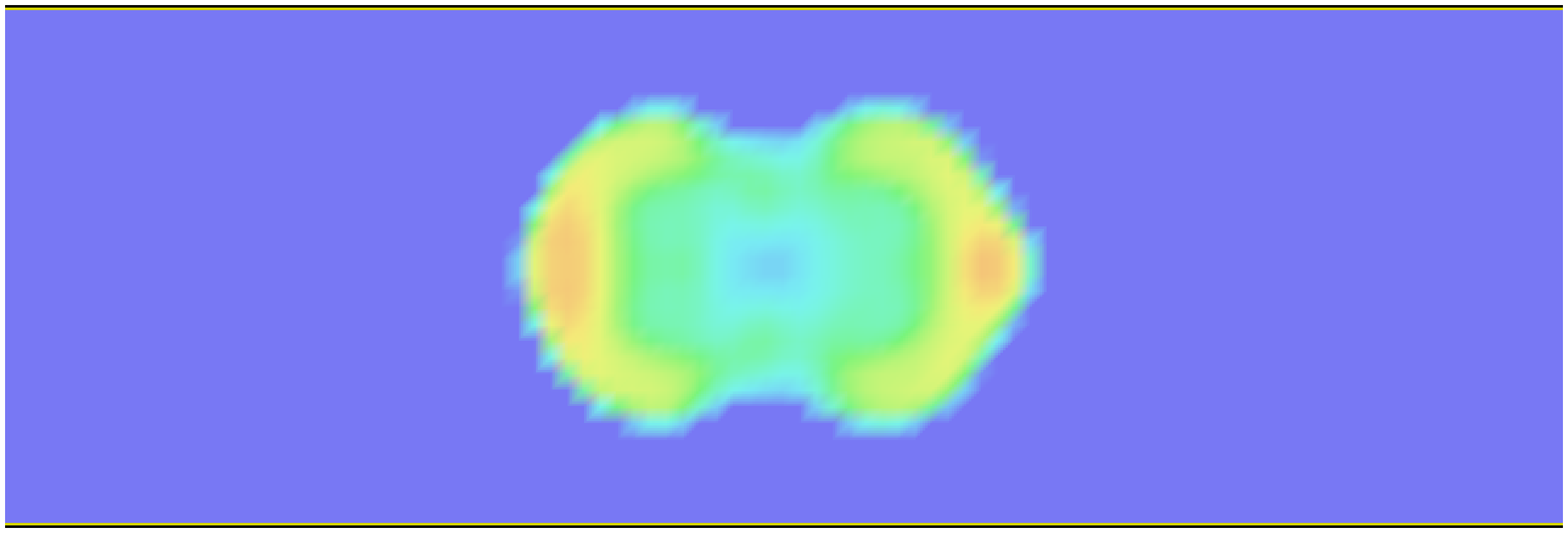}}\\
\subfloat{\includegraphics[width=0.05\textwidth, angle=90]{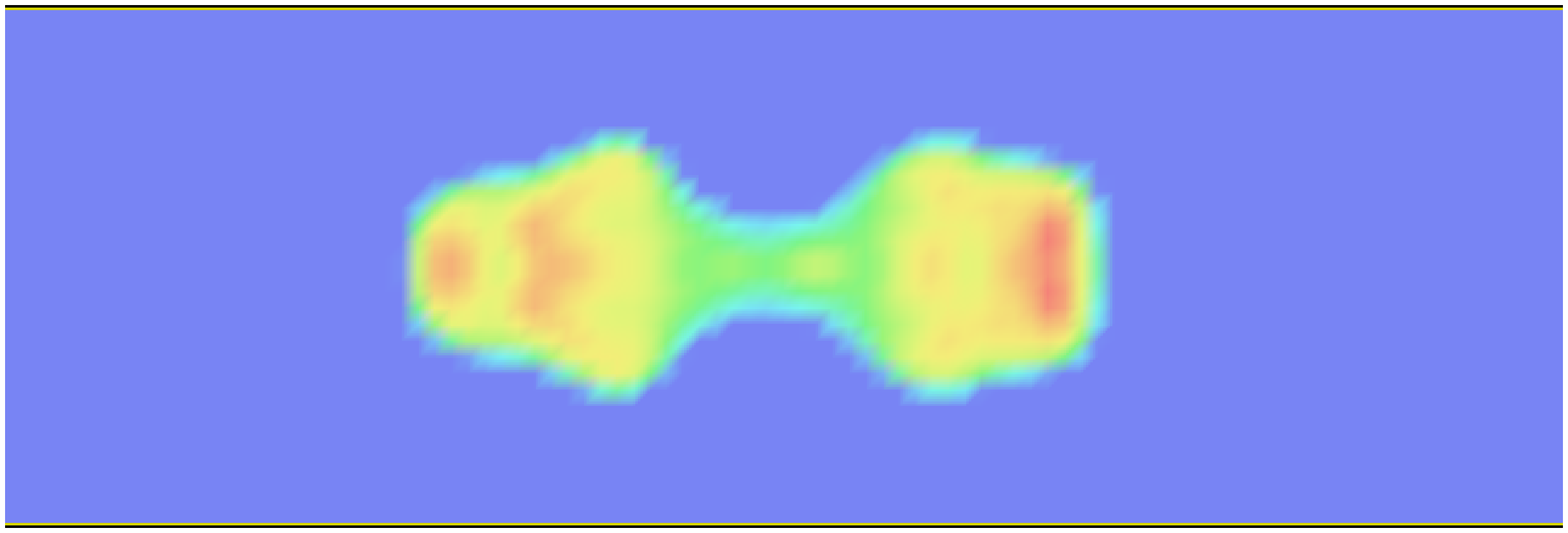}}
\subfloat{\includegraphics[width=0.05\textwidth, angle=90]{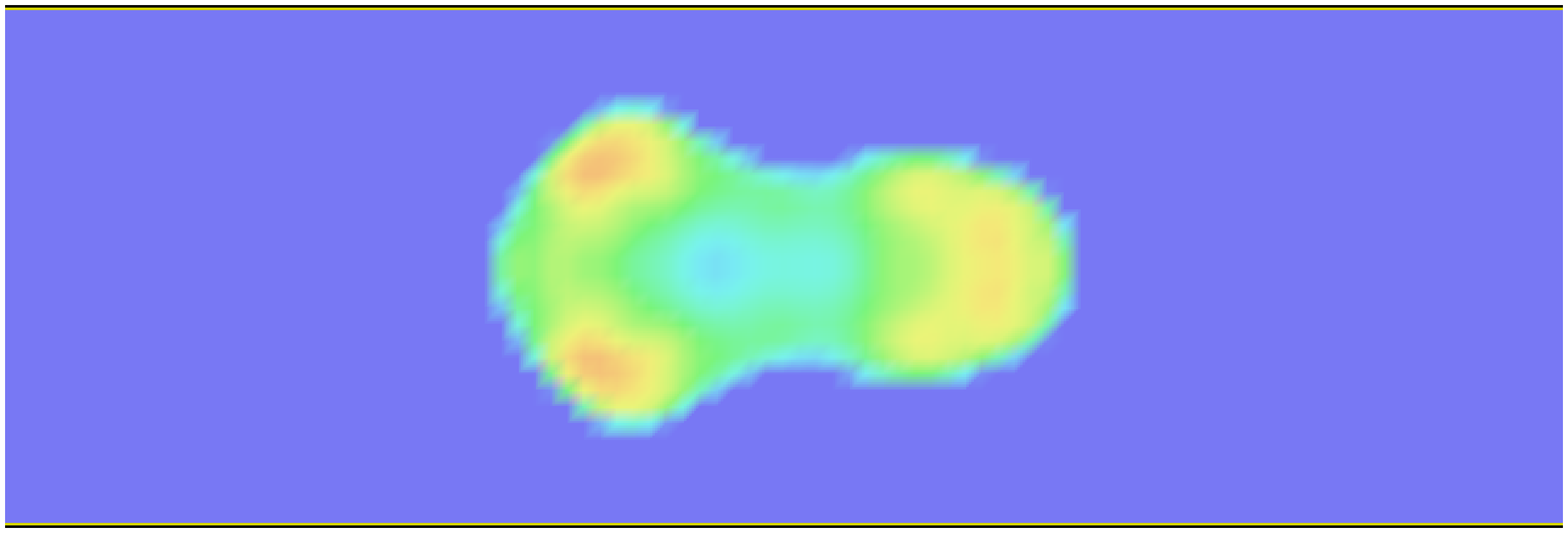}}
\subfloat{\includegraphics[width=0.05\textwidth, angle=90]{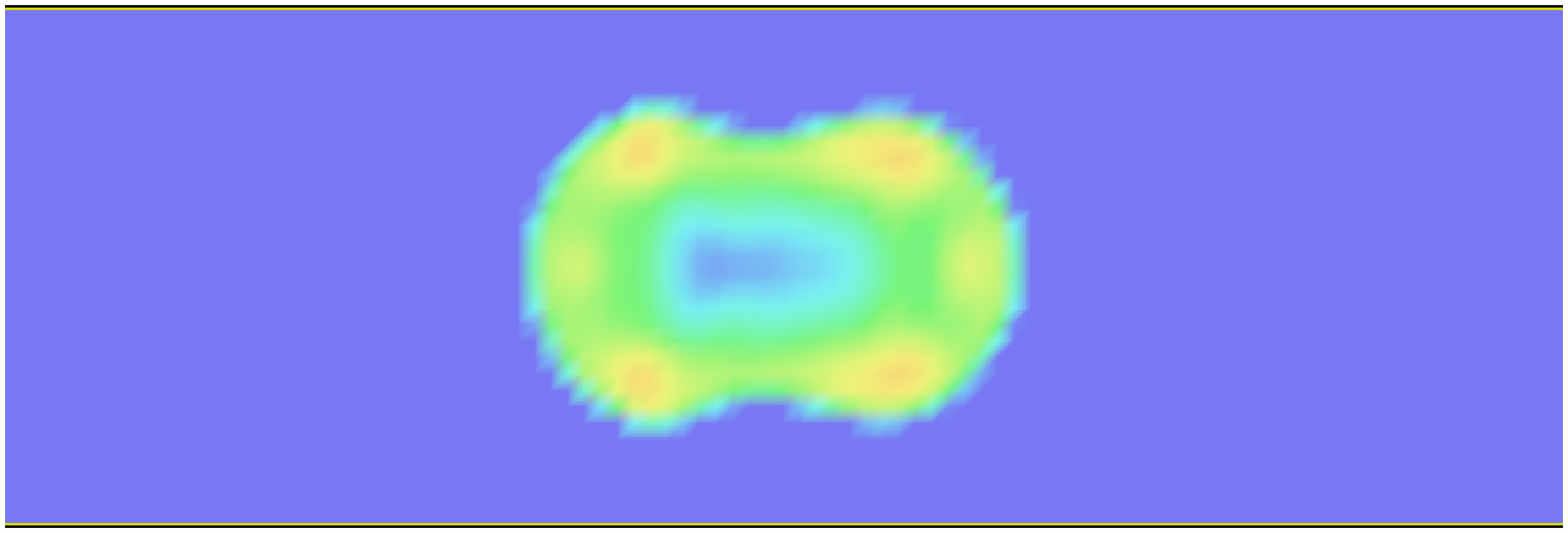}}\\
\subfloat{\includegraphics[width=0.05\textwidth, angle=90]{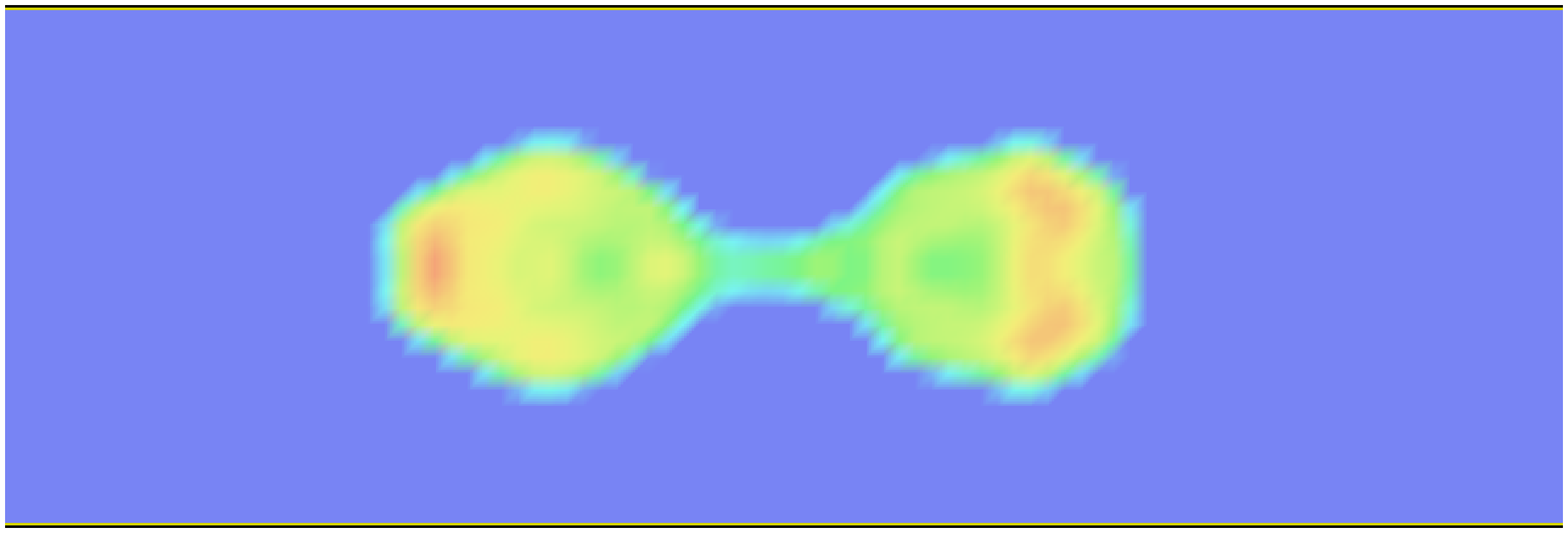}}
\subfloat{\includegraphics[width=0.05\textwidth, angle=90]{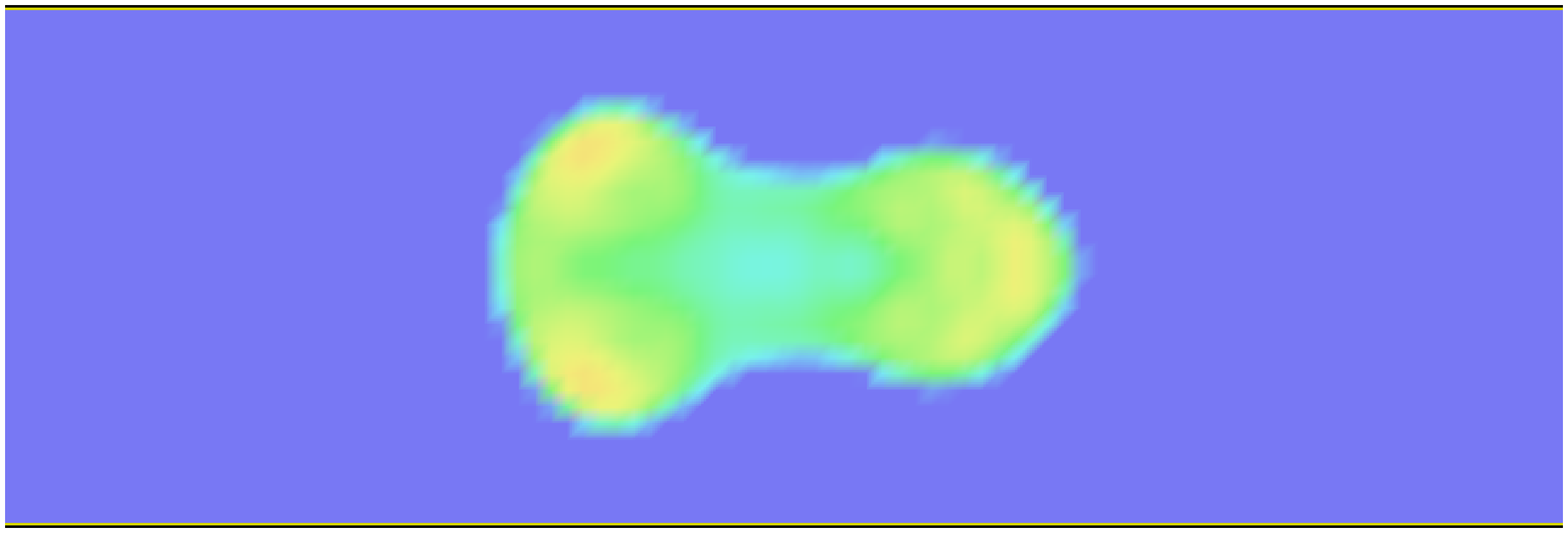}}
\subfloat{\includegraphics[width=0.05\textwidth, angle=90]{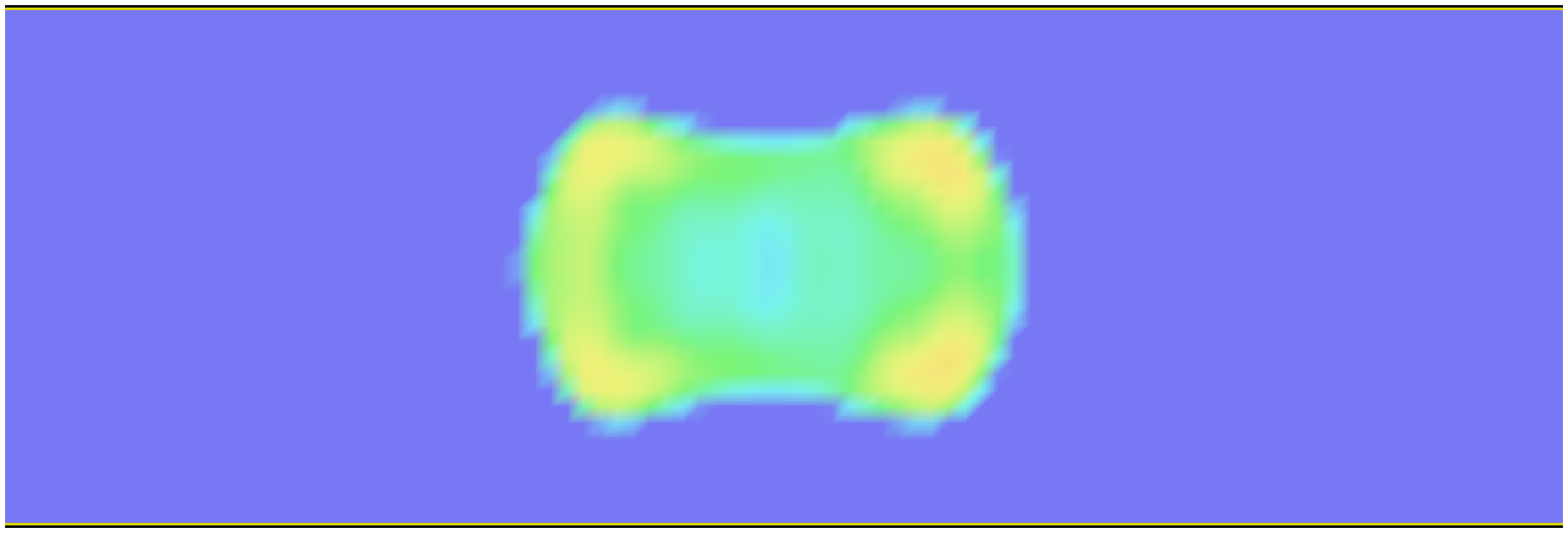}}\\
\subfloat{\includegraphics[width=0.05\textwidth, angle=90]{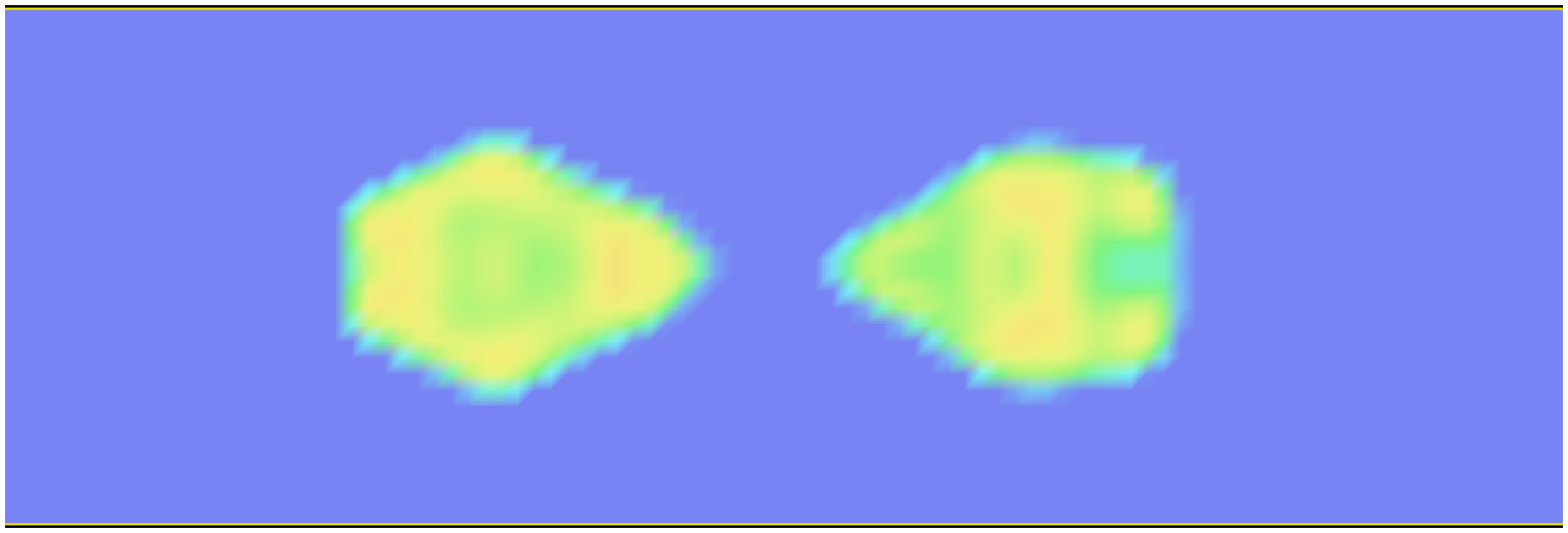}}
\subfloat{\includegraphics[width=0.05\textwidth, angle=90]{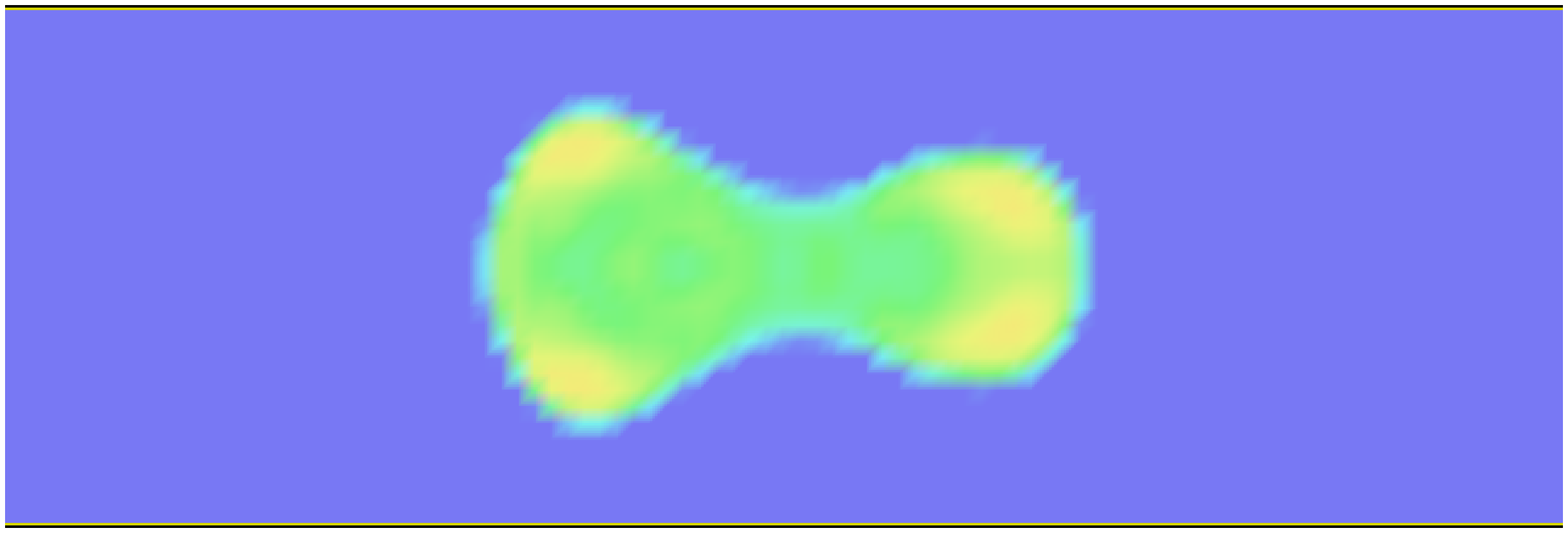}}
\subfloat{\includegraphics[width=0.05\textwidth, angle=90]{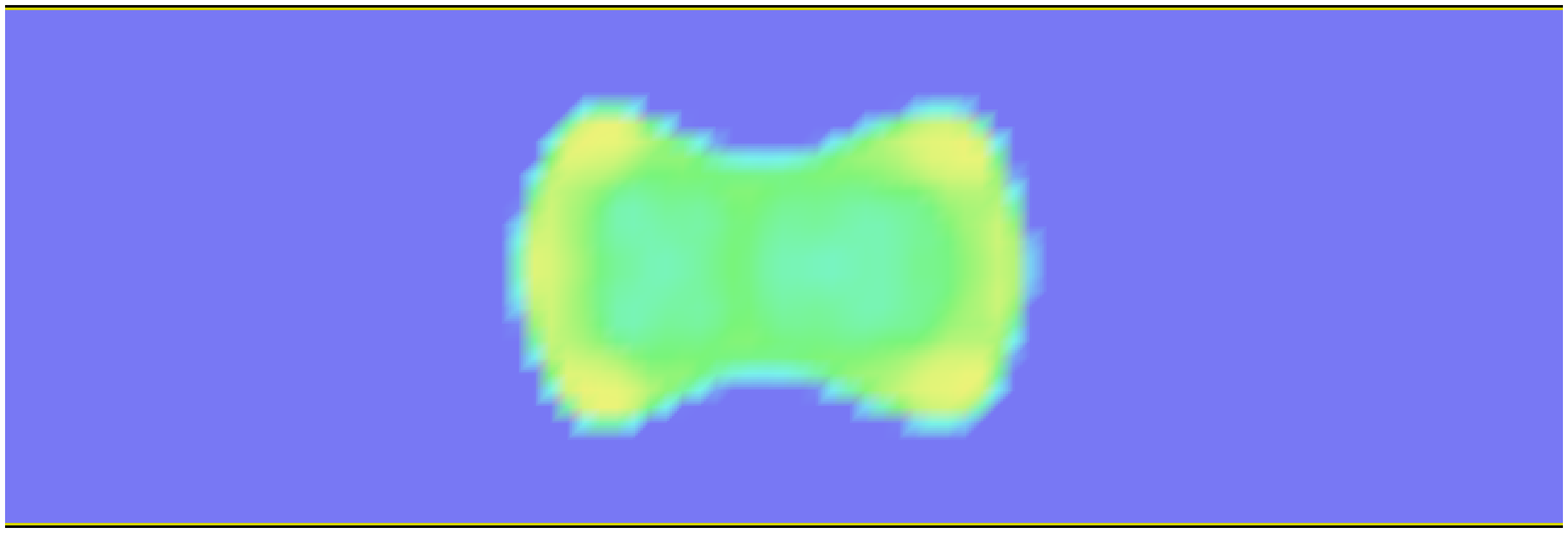}}\\
\subfloat{\includegraphics[width=0.05\textwidth, angle=90]{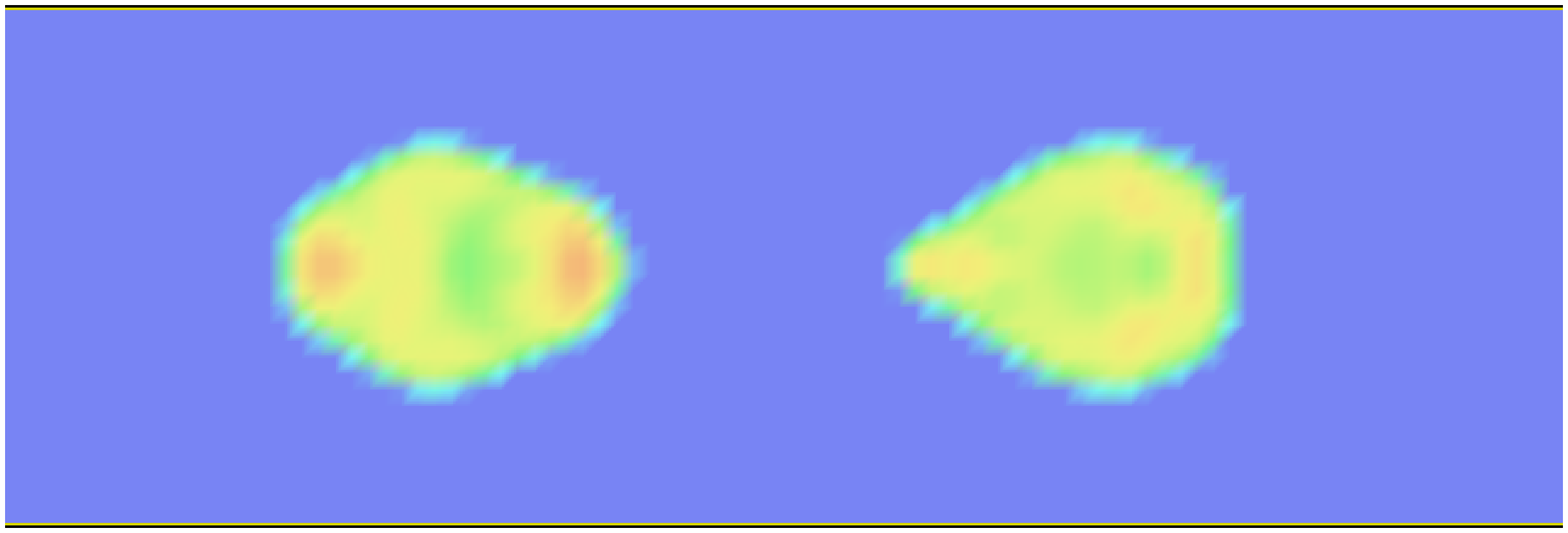}}
\subfloat{\includegraphics[width=0.05\textwidth, angle=90]{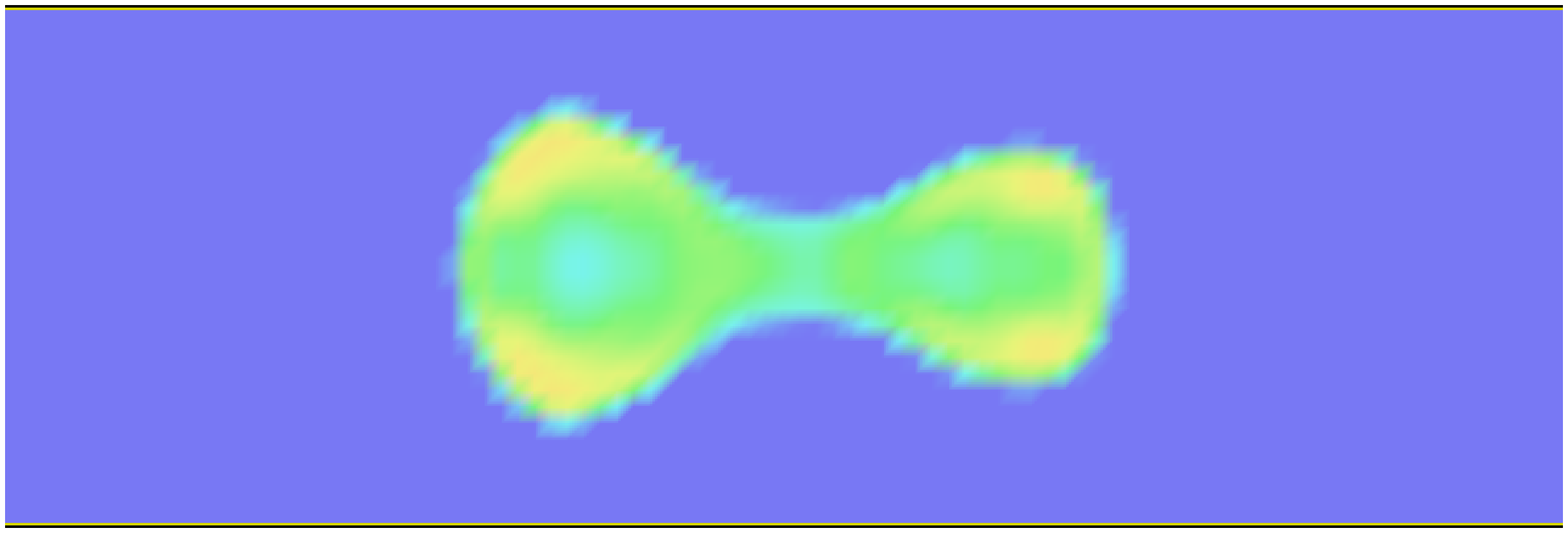}}
\subfloat{\includegraphics[width=0.05\textwidth, angle=90]{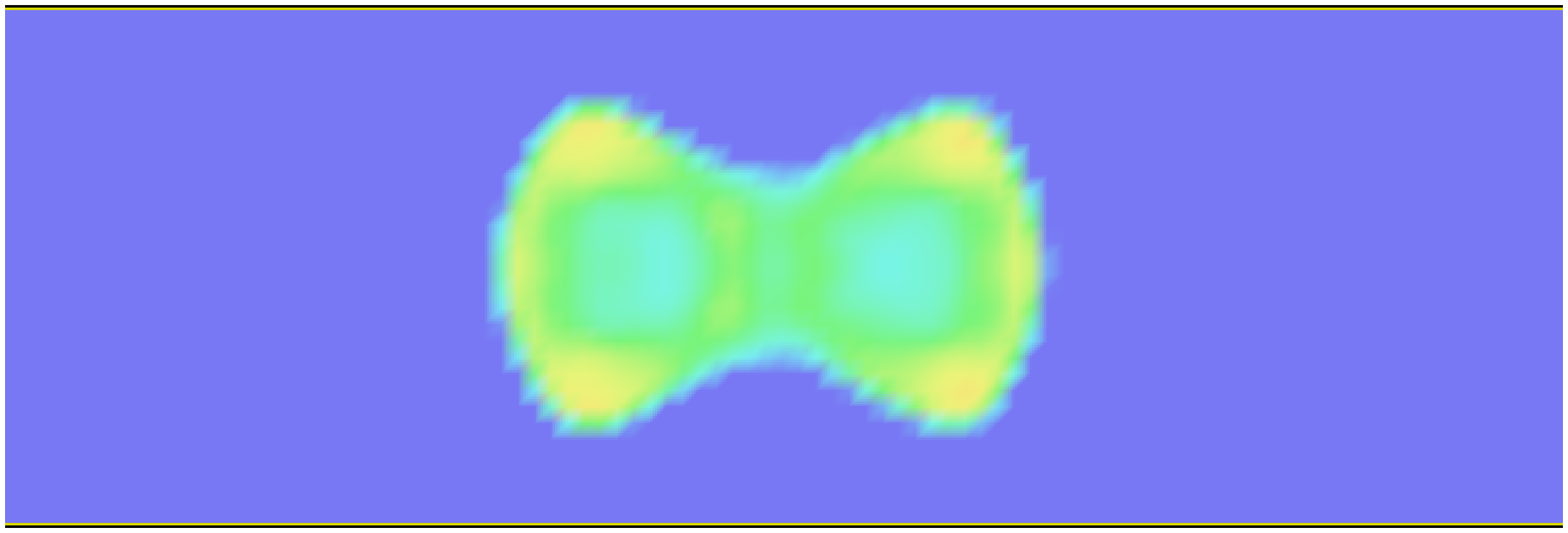}}\\
\subfloat{\includegraphics[width=0.05\textwidth, angle=90]{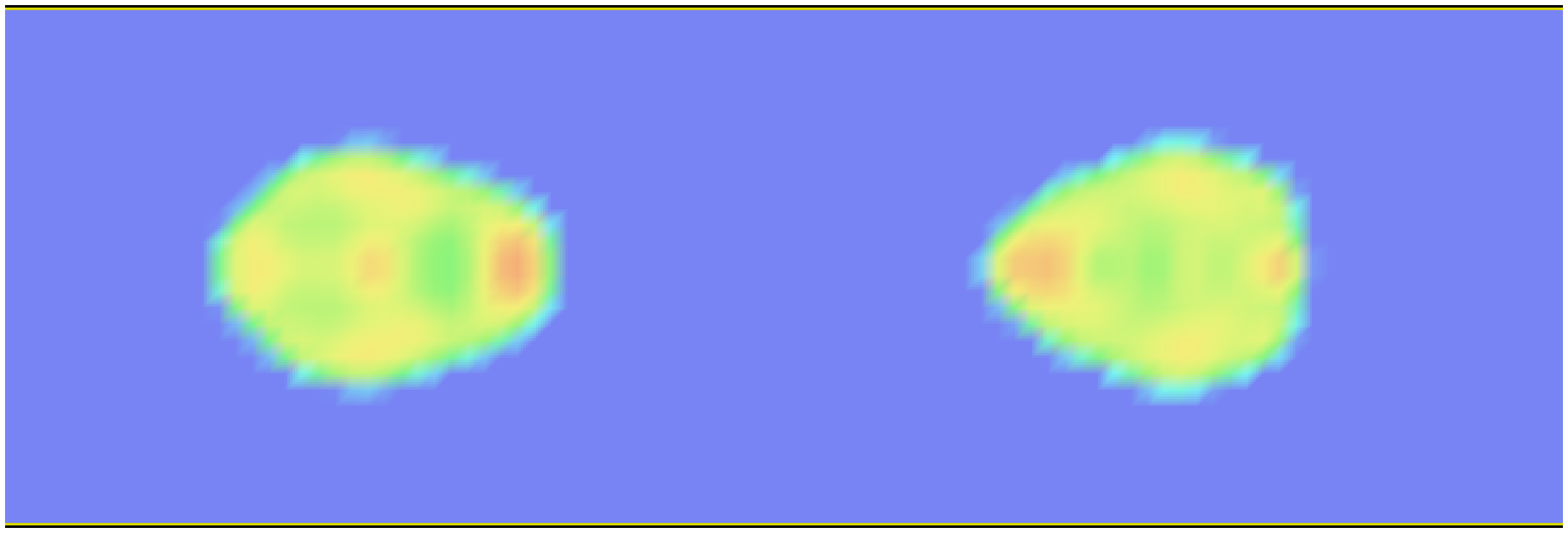}}
\subfloat{\includegraphics[width=0.05\textwidth, angle=90]{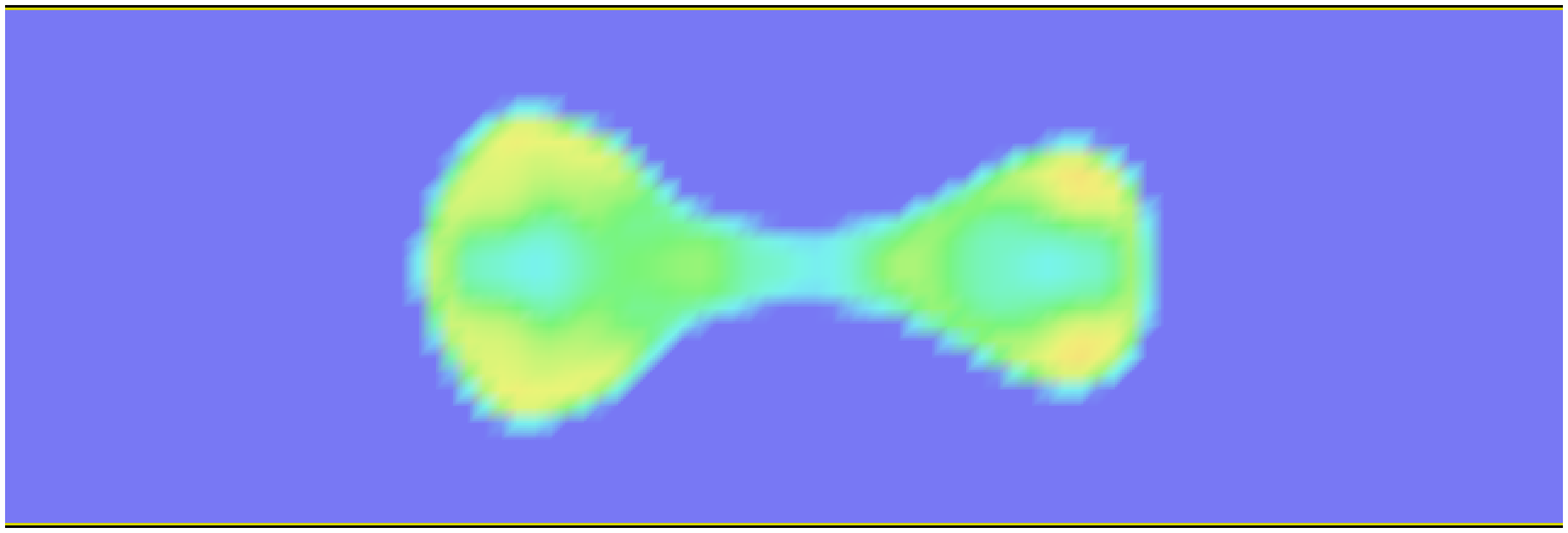}}
\subfloat{\includegraphics[width=0.05\textwidth, angle=90]{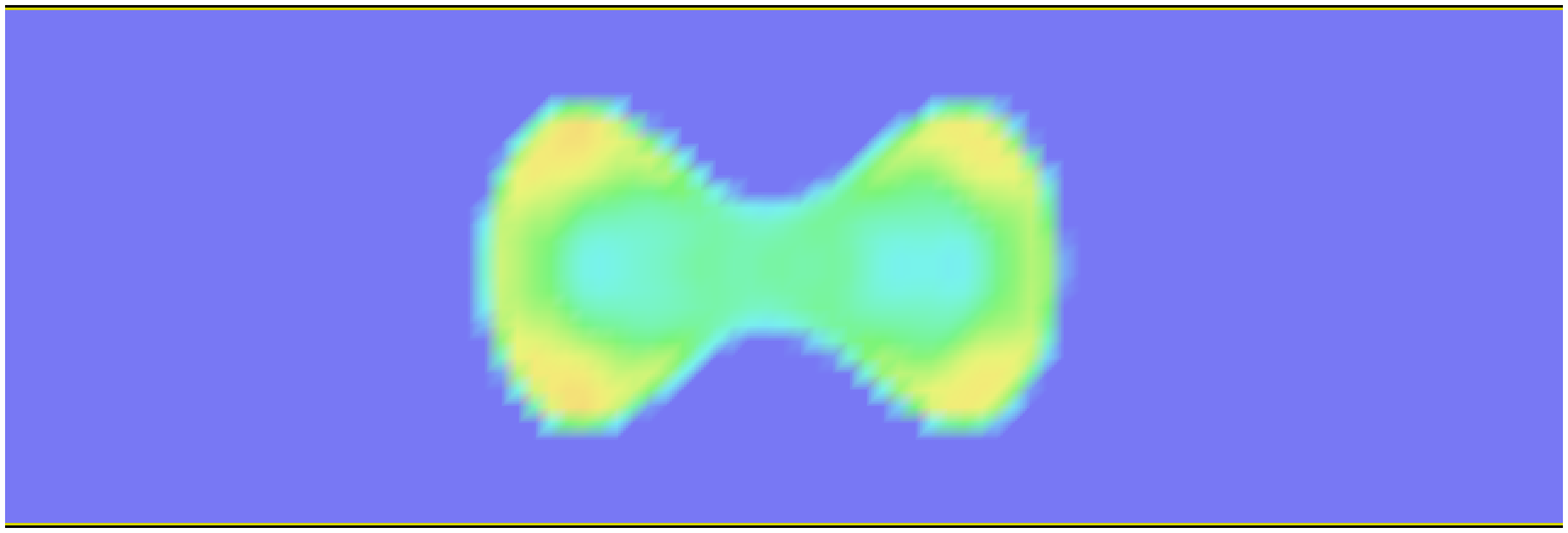}}\\
\subfloat{\includegraphics[width=0.05\textwidth, angle=90]{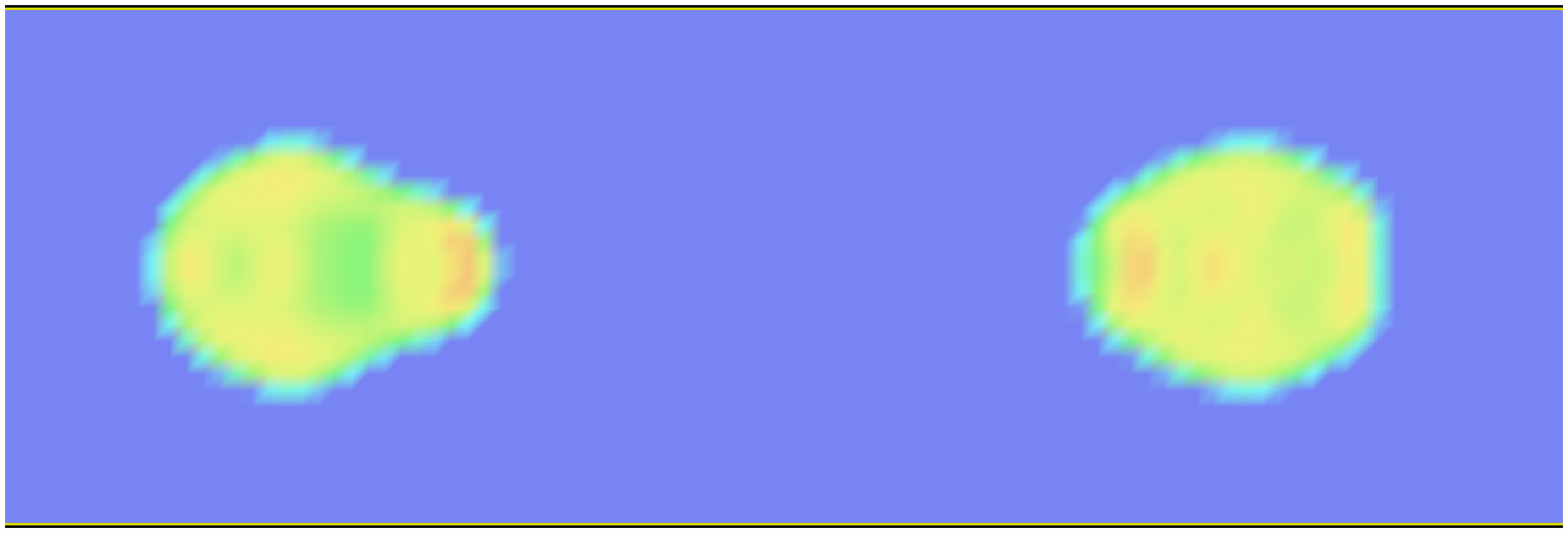}}
\subfloat{\includegraphics[width=0.05\textwidth, angle=90]{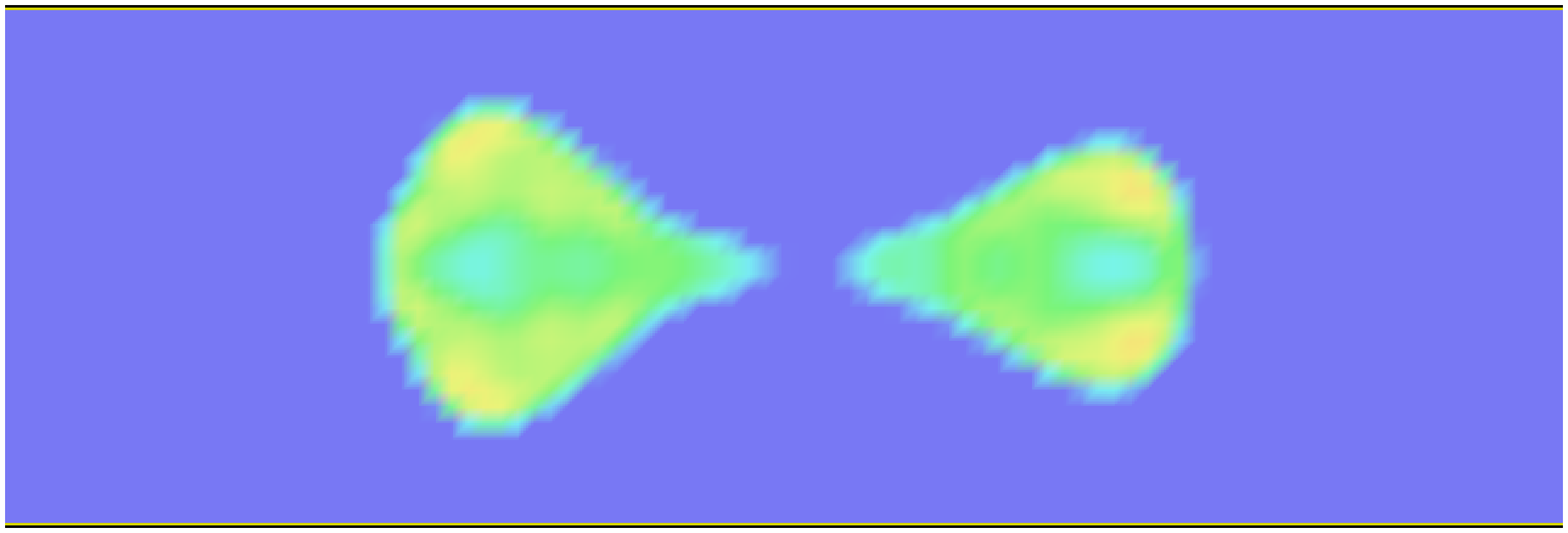}}
\subfloat{\includegraphics[width=0.05\textwidth, angle=90]{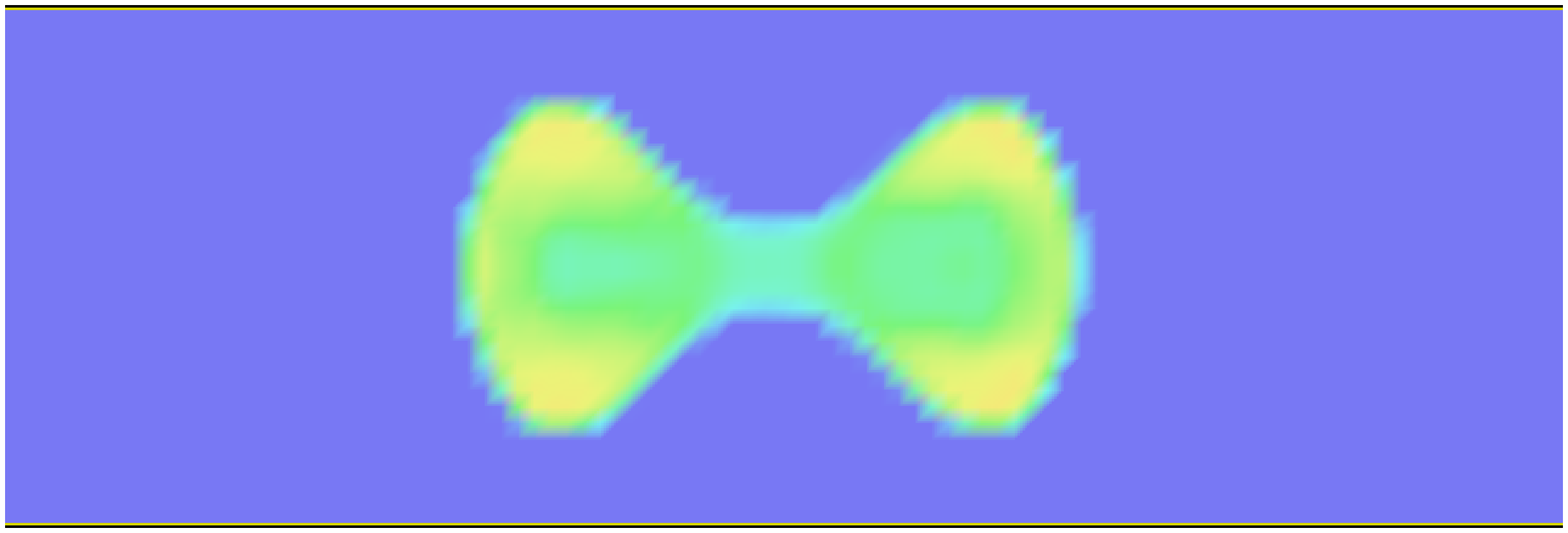}}
\end{center}
\caption{\label{Fig:Snapshots}(color online). Nucleon density in the $z=0$ plane at various times for a central collision of a $^{250}$Cf (initially on the left) with a $^{232}$Th (right) nucleus at a centre of mass energy $E_{{c.m.}}=1012.2$~MeV. Snapshots are shown from $t=2.4\times10^{-22}$~s to $3.54\times10^{-21}$~s in time steps of $3\times10^{-22}$~s from top to bottom. From left to right, the columns represent the XX, YX and YY  relative orientations (see text). Dark blue denotes densities below $0.1$~fm$^{-3}$ and dark red marks those above $0.16$~fm$^{-3}$.}
\end{figure}

Both nuclei exhibit a strong prolate deformation in their ground state and can, 
in principle, take all possible orientation in the entrance channel. 
Five different relative orientations between the nuclei, labelled XX, XY, YX, YY, and YZ, 
have been selected to study their role on the reaction mechanism (see top of Fig.~\ref{Fig:Snapshots} and Fig.~\ref{Fig:legende}). 
We define them according to how the elongation axes are angled with the collision axis (i.e., the $x$~axis). 
For instance, the XX (YY) orientation involves the two nuclei colliding on their tips (sides), 
as shown by the left (right) column of Fig.~\ref{Fig:Snapshots}. 
In the XY and YX orientations, contact occurs first between  
the tip of one nucleus and the side of the other.
The first letter corresponds to the heavier nucleus. 
Thus, the central column of  Fig.~\ref{Fig:Snapshots} displays the YX orientation, 
with the elongation axis of $^{250}$Cf  ($^{232}$Th) perpendicular (parallel) to the collision axis. 
Figure~\ref{Fig:Snapshots} clearly shows the importance of the initial orientation on the reaction mechanism.
For instance,  in the last snapshot, the fragments in the XX configuration are well separated while 
a neck is still present in the other orientations, showing a shorter contact time in the XX case. 
The internal density and shape evolutions also depends on the orientation, 
going from strong fluctuations in the XX orientation to a smooth evolution in the YY one.
Finally, the YX orientation produces the heaviest element (left fragment), corresponding to a transfermium nucleus.

\subsection{Multinucleon transfer in central collisions}
\label{subsec:multinucleon}

The process of standard quasifission is usually dominant in reactions with heavy nuclei
where nucleons are transferred from the heavier to the lighter nucleus. 
As the dinuclear system is electrostatically unstable, 
it then separates into two fragments, with an increased mass symmetry.
The production of transfermium nuclei in  $^{232}$Th+$^{250}$Cf
implies that a product nucleus has to have more mass than either of the original two.
Quasifission must either act in reverse, with nucleon transfer from the lighter to the heavier nucleus, called inverse quasifission (IQ),
or overshoots so that $^{232}$Th attains enough nucleons to end up heavier than $^{250}$Cf, which we define as swap-IQ.

To count the number of protons, $Z_{f_i}$, and neutrons, $N_{f_i}$, in the fragment $i$ 
($i=1,2$ as all the present calculations show only two fragments in the exit channel), 
an integration of the corresponding densities is performed 
in space regions where the total density exceeds $0.001$~fm$^{-3}$ at the last iteration time.
Applying this procedure at the initial time step excludes approximately $0.7$ bound neutrons per fragment and no proton. 
The remnant neutrons are found in the tails of the wave functions.
Thus this value is added to the integration of density to obtain a correct estimate of $N_{f_i}$.
Note that the TDHF evolution is stopped before the fragments reach the walls of the box 
to avoid nucleon emission due to unrealistic rebounds.

\begin{figure}[h]
\begin{center}
\subfloat[\label{Fig:legende}]{\includegraphics[width=0.4\textwidth]{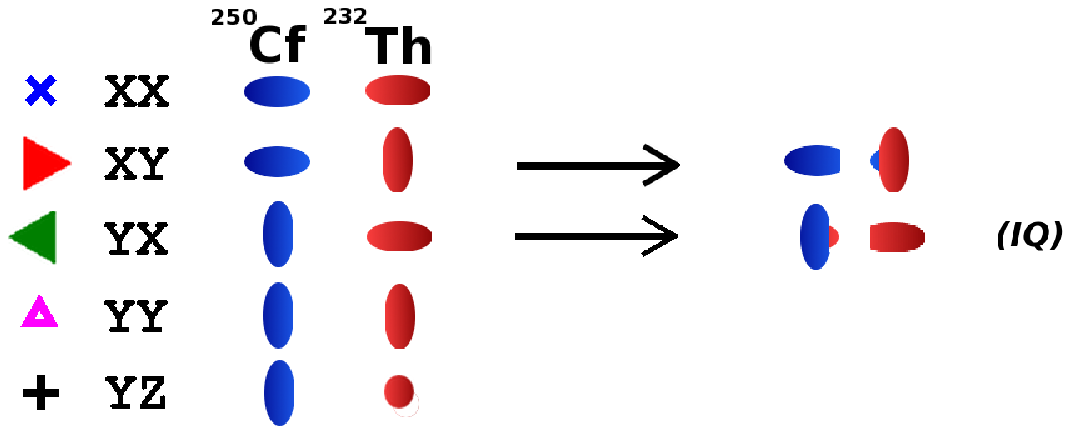}}\\
\subfloat[\label{Fig:ProtonChange}]{\includegraphics[width=0.42\textwidth]{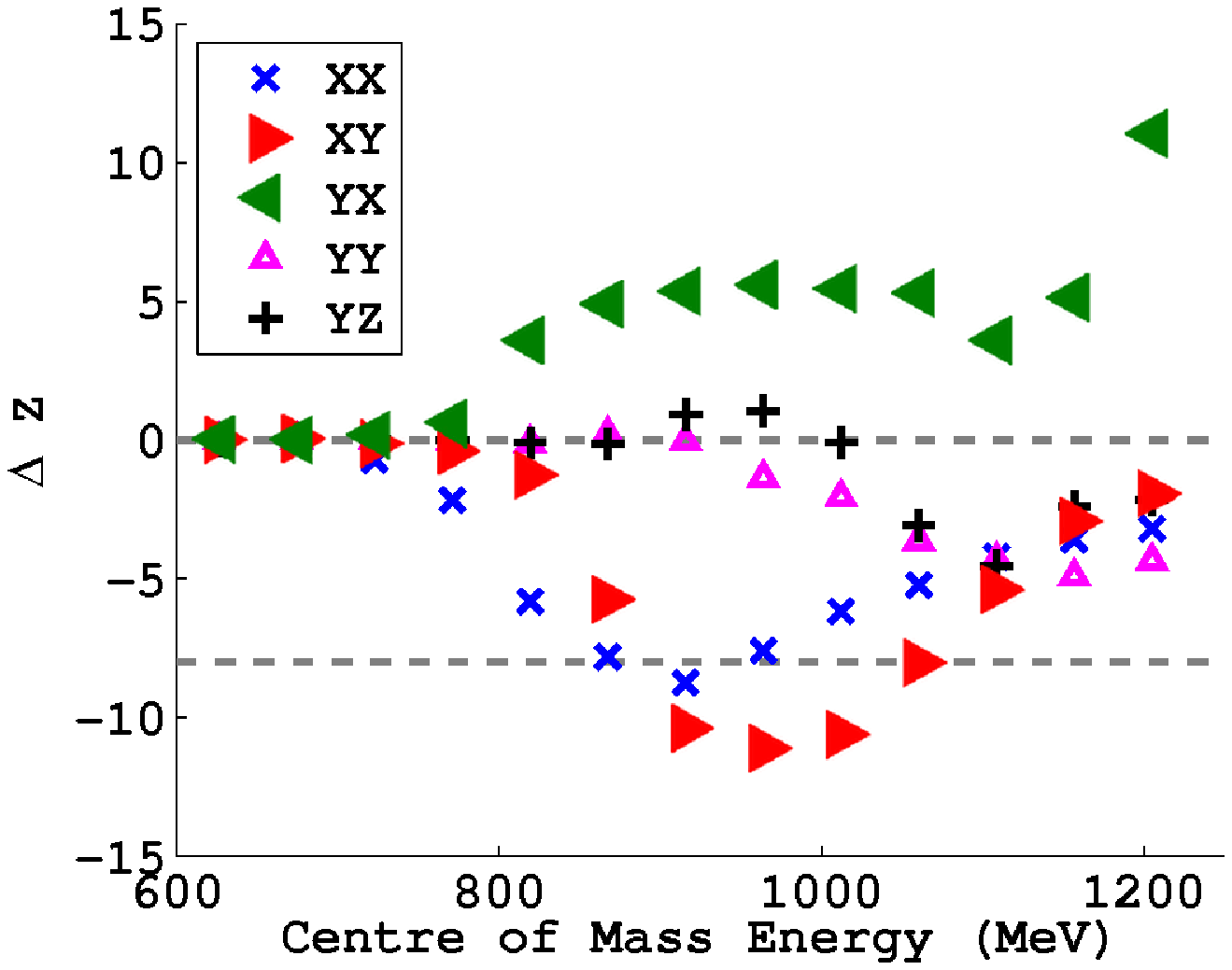}}\\
\subfloat[\label{Fig:NeutronChange}]{\includegraphics[width=0.42\textwidth]{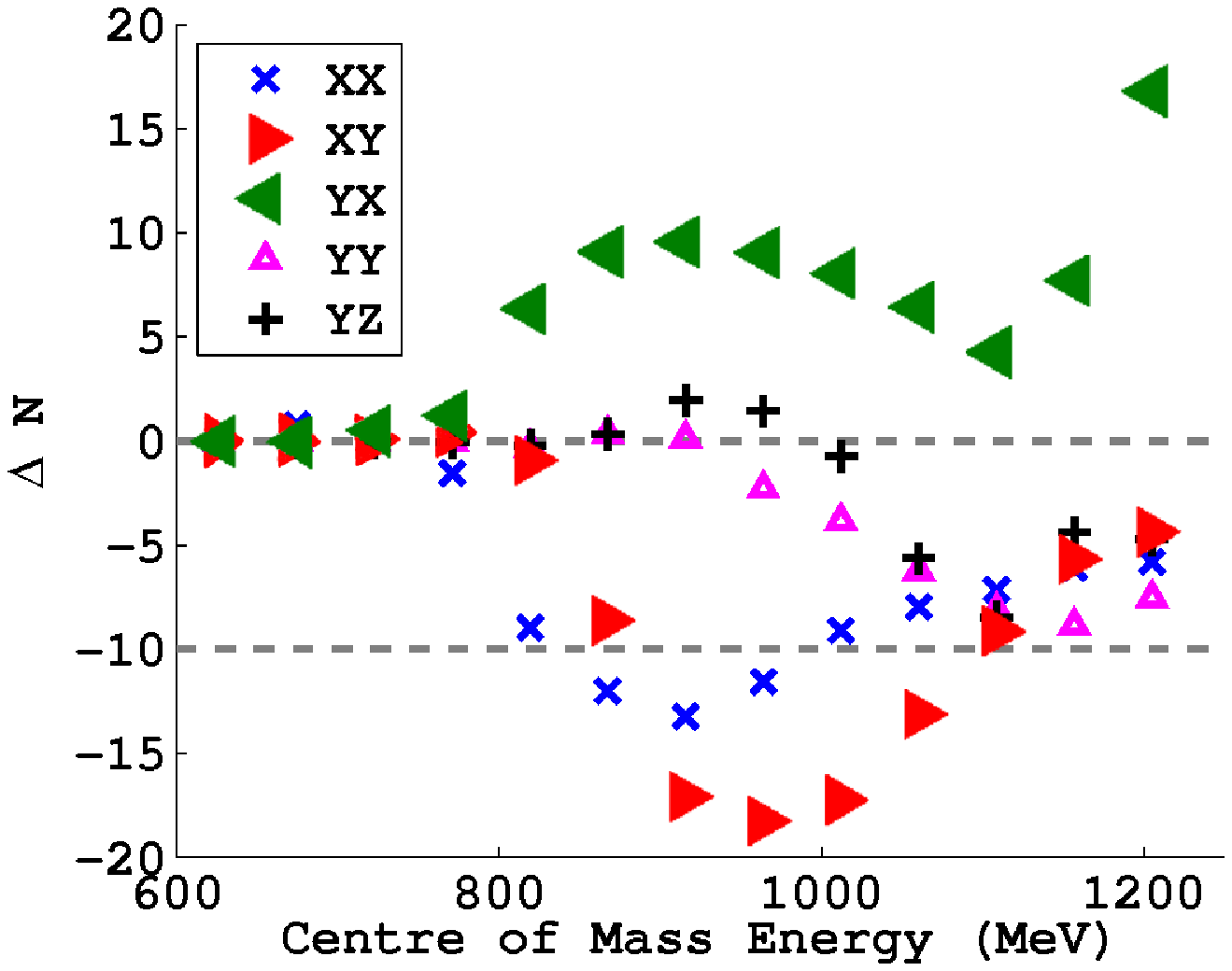}}
\end{center}
\caption{\label{Fig:NucleonChange}(color online). (a) Representation of the initial relative orientations and sketch of the transfer from the tip to the side in the XY and YX configurations. Variation of (b) proton and (c) neutron numbers in the $^{250}$Cf-like fragment after collision with a $^{232}$Th nucleus at varying centre of mass energies for five relative orientations. 
The dashed lines at $0$ represent $^{250}$Cf and the lines at $\Delta {Z} = -8$ and $\Delta {N} = -10$ mark $^{232}$Th.}
\end{figure}

The dependence on beam energy of the proton and neutron numbers in the $^{250}$Cf-like fragment 
are plotted in Figs.~\ref{Fig:ProtonChange} and~\ref{Fig:NeutronChange}, respectively, for each initial relative orientation. 
The dashed lines represent the change in nucleon value required to end either as a $^{250}$Cf or a $^{232}$Th in the exit channel.
Events lying between these two lines correspond to standard quasifission, 
while events above the upper and below the lower line are associated with IQ and swap-IQ processes, respectively.
While most events are located around or between these lines, the YX configuration leads clearly to a strong IQ 
for centre of mass energies $E_{{c.m.}}>800$~MeV.
Here, the tip of $^{232}$Th comes into contact with the side of $^{250}$Cf and is absorbed (see the sketch of the exit channel in Fig.\ref{Fig:legende}). 
Note that the rapid increase of the number of transfered nucleons around $E_{{c.m.}}\simeq1200$~MeV
can be attributed to strong dynamical fluctuations of the internal density, modifying the breaking point of the dinuclear system (see Ref.~\cite{gol09} and section~\ref{subsec:time}), rather than standard transfer where the flux of nucleons occurs with a smooth change of the shape of the fragments.

Focusing on the low-energy range ($E_{c.m.}<1000$~MeV) of the YX orientation in Fig.~\ref{Fig:NucleonChange}, 
the most massive nucleus, corresponding to $^{265}$Lr, is formed at $915.8$ MeV. 
This corresponds to three neutrons heavier than the heaviest Lawrencium isotope found experimentally to date~\cite{blo10}. Furthermore, this corresponds only to the expected center of the fragment mass and charge distributions for this particular orientation.
Taking into account particle number fluctuations in the fragment (which are known to be underestimated in TDHF~\cite{das79}) would lead to more neutron-rich nuclei in the tail of the fragment mass distribution. 
Finaly, let us note that heavy fragments are also produced by swap-IQ in the XY orientation at 
$E_{{c.m.}}\simeq950$~MeV.

\subsection{Fast neutron evaporation}

\begin{figure}[h]
\includegraphics[width=0.42\textwidth]{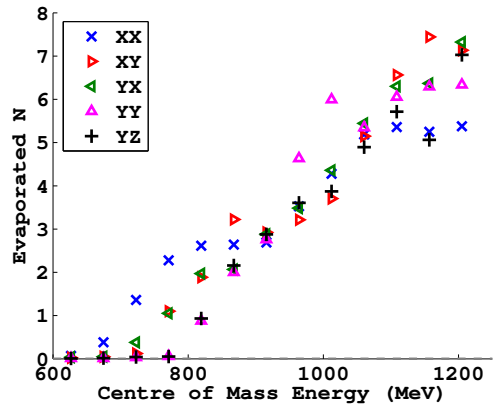}
\caption{\label{Fig:NeutronEvap}(color online). The total number of neutrons evaporated from the $^{250}$Cf and $^{232}$Th fragments for varying centre of mass collision energies and five different relative orientations.}
\end{figure}

During the collision, a significant part of the relative kinetic energy of the nuclei 
is transformed into internal excitation of the dinuclear system and its fragments.
Thus, the system can emit particles, in particular neutrons, before and after its separation~\cite{hin92}.
On one hand, this neutron emission reduces the chance to produce neutron-rich nuclei, 
but, in the other hand, it is a cooling mechanism which increases the survival probability of the fragments against secondary fission.

In principle, nucleon emission is a one-body process accounted for in TDHF.
However, the finite time of the calculation allows only to estimate the total number 
of emitted nucleons soon after the re-separation of the fragments (typically about $10^{-21}$~s after the neck breaks).  
To get a quantitative insight into neutron emission, 
the total number of neutrons lost to the fragments is computed at the end of the calculation. 
As the TDHF evolution is unitary and conserves the total number of neutrons, $N_{tot.}=294$, 
the number of evaporated neutrons is determined from the relation $N_{evap.}=N_{tot.}-N_{f_1}-N_{f_2}$.
Figure~\ref{Fig:NeutronEvap} gives the evolution of $N_{evap.}$ for various orientations and energies.
A global linear increase of emitted neutrons is observed with energy. 
In contrast, the calculations shows that no proton has been lost to either fragment, 
due to the Coulomb barrier at the surface of the nuclei. 


The way the number of nucleons in the fragments is defined in section~\ref{subsec:multinucleon}
implies that the variations of the neutron number in the $^{250}$Cf-like fragment in Fig.~\ref{Fig:NeutronChange}
already takes into account this neutron emission. 
For instance, in the YX case at $E_{c.m.}=915.8$~MeV, leading to the $^{265}$Lr nucleus, 
approximately three neutrons have been evaporated, i.e., $\sim1-2$ neutron per fragment,
carrying away some of their excitation energy.
The subsequent decay occurs by neutron and gamma emission, and by secondary fission.
{ The question of the remaining excitation energy of the fragments 
is essential to determine their survival probability against fission in one hand, 
and, in the other hand, to predict which isotopes are finally produced.
For instance, for all IQ events in the YX configuration, the final total kinetic 
energy of the fragments predicted by TDHF is of the order of $\sim650$~MeV.
Then, to enhance the survival probability of neutron-rich heavy nuclei,
it may be preferable to consider a lower energy than the optimum one 
deduced by Fig.~\ref{Fig:NucleonChange}~\cite{hei10}.}

\subsection{Collision time and saturation in the neck}
\label{subsec:time}

The multinucleon transfer is expected to be affected by the life-time of the dinuclear system,
i.e., the time during which the two fragments are in contact.
The collision time is also an important input for example in the calculations of electron-positron pair production 
from the quantum electrodynamics (QED) vacuum decay~\cite{rei81,gre83,ack08}.
As for multinucleon transfer, the collision time between actinides has been recently investigated 
in various models~\cite{mar02,tia08,sar09,zha09,gol09,zag06,zag07} as well as experimentally~\cite{gol10}.
In particular, it has been shown that the collision time depends 
on the initial orientation~\cite{gol09}.
Indeed, as one can see in Fig.~\ref{Fig:Snapshots}, it is much smaller for the XX orientation. 

\begin{figure}[h]
\includegraphics[width=0.42\textwidth]{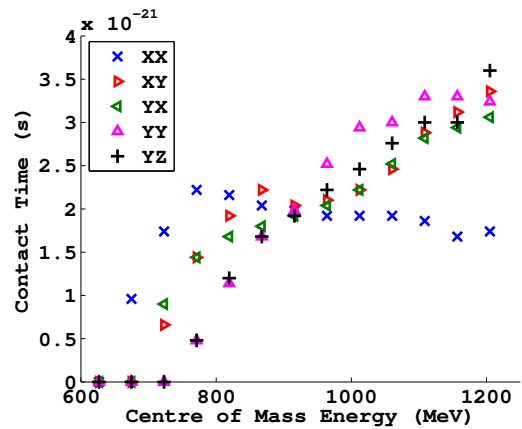}
\caption{\label{Fig:ContactTime}(color online). Time during which the fragments are in contact, 
for five relative orientations, as function of the center of mass energy.}
\end{figure}

Following Ref.~\cite{gol09}, we define the collision (or ''contact'') time as the time during which
the fragments are in contact with a neck density exceeding one tenth of the saturation density, i.e., $\rho_{neck}\ge\rho_0/10=0.016$~fm$^{-3}$.
Figure~\ref{Fig:ContactTime} presents the evolution of this time with energy. 
The same behavior as in the $^{238}$U+$^{238}$U case is observed in the present energy range 
(see Fig.~2b of Ref.~\cite{gol09}), i.e.,
an increase with energy up to $3-4\times10^{-21}$~s at $E_{c.m.}=1200$~MeV 
for all orientations except the XX one which exhibits a plateau at $\sim2\times10^{-21}$~s. 
Comparing Figs.~\ref{Fig:NucleonChange} and~\ref{Fig:ContactTime}, it is interesting to note that
the absolute value of the number of transfered nucleons and the contact times have very different behaviors.
This may be attributed to the dynamics of the dinuclear system, 
in particular in term of its complex shape evolution (see, e.g., Fig.~\ref{Fig:Snapshots}).
Note that the decrease of the collision time at higher energy observed in~\cite{gol09} 
is outside the energy range of the present calculations.

\begin{figure}[h]
\includegraphics[width=0.42\textwidth]{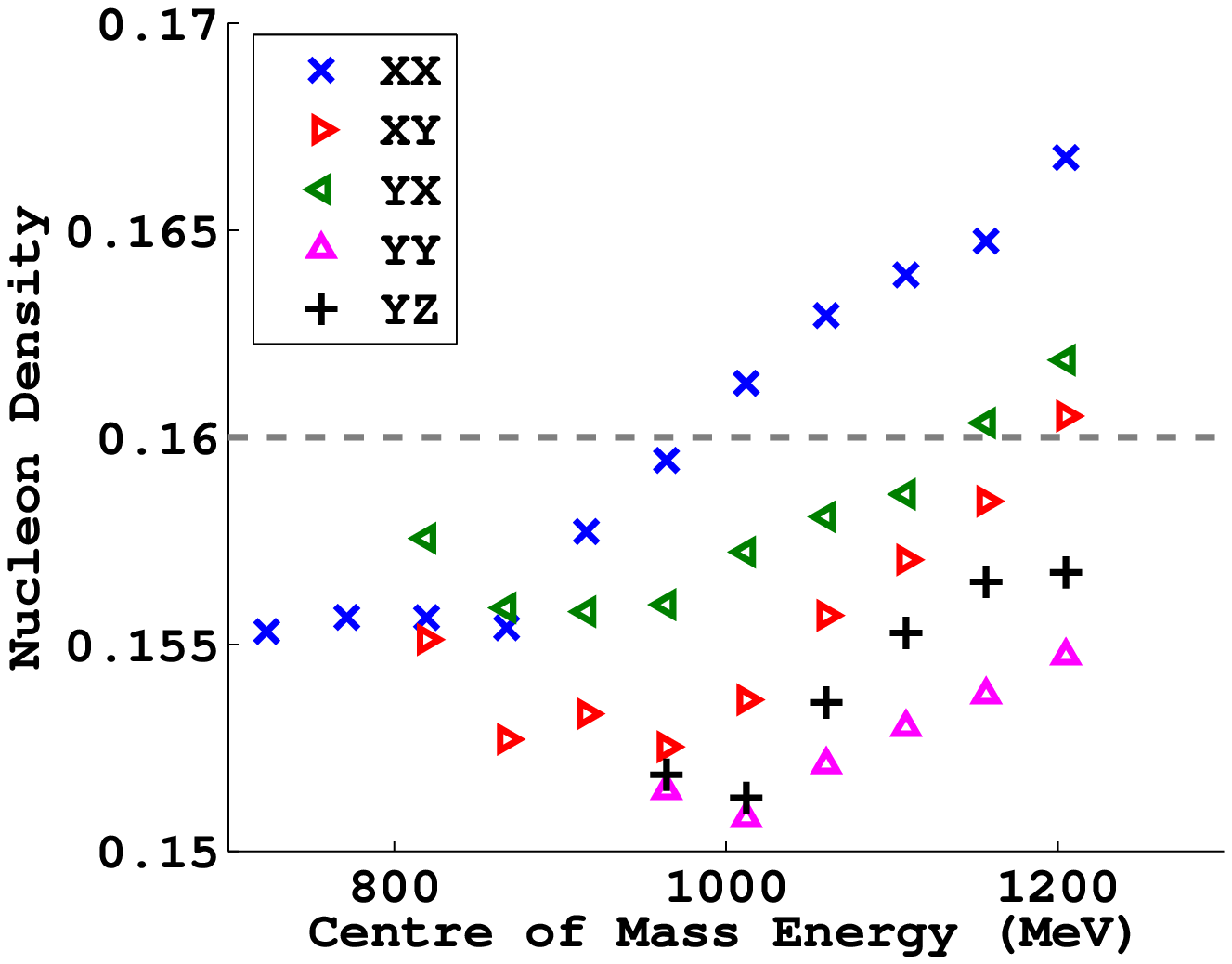}
\caption{\label{Fig:MaxOverlapDensity}(color online). Maximum density along the collision axis between $^{250}$Cf and $^{232}$Th, for varying centre of mass energies and five relative orientations. The dashed line represents the saturation density $\rho_0=0.16$~fm$^{-3}$ and points are displayed only if the two nuclei overlap, as decided by a minimum central axis density greater than $0.14$ fm$^{-3}$.}
\end{figure}

As in the uranium-uranium case, the saturation of the collision time in the XX orientation 
can be interpreted as an effect of the overcoming of the saturation density in the neck,
inducing a strong repulsion between the fragments.
To get a deeper insight into this effect, let us study the maximum density
for two overlapping nuclei. The criterion to define that the nuclei overlap 
is that the minimum density in the neck region 
on the collision axis has to be greater than $0.14$~fm$^{-3}$.
Fig.~\ref{Fig:MaxOverlapDensity} shows the maximum density along the collision axis during the overlap.
The maximum density increases with energy above an energy threshold 
which depends on the initial orientation of the nuclei.
A nucleus with an orientation X, i.e., with its deformation axis along the collision axis,
overlaps with its collision partner sooner, and at lower energy than for the other orientations.
This is why the energy threshold $E_{th.}$ above which the maximum density increases is lower for XX ($E_{th.}^{XX}\simeq870$~MeV) than for YX and XY ($E_{th.}^{YX,XY}\simeq960$~MeV),
which, in turn, are also lower than for YY and YZ ($E_{th.}^{YY,YZ}\simeq1010$~MeV).
As a consequence, the maximum density exceeds the saturation density in the XX at lower energy 
(typically for $E\ge1000$~MeV), than in the other orientations. 
It is interesting to note, however, that the plateau in the contact time in the XX orientation (see Fig.~\ref{Fig:ContactTime})
starts at lower energy (at about $\sim770$~MeV) than $E_{th.}^{XX}$ and that other dynamical effects 
may also play a role in the saturation of the contact time.

We also note that the sudden increase of IQ nucleon transfer for orientation YX at $E_{{c.m.}}\simeq1200$~MeV 
in Fig.~\ref{Fig:NucleonChange} coincides with densities surpassing saturation in Fig.~\ref{Fig:MaxOverlapDensity}. 
A close look at the internal density evolution shows that this overcoming of $\rho_0$ generates fast dynamical fluctuations.
As a consequence, the system can break at different positions, inducing variations of the number of nucleons in the fragments
that are not totally due to standard multinucleon transfer through the neck.
It is important to note that this effect occurs only in violent collisions, and that the excited fragments 
should have a very small chance to survive against subsequent fission.

\subsection{Role of impact parameter}

We finally investigate how multinucleon transfer evolves with impact parameter
 for the YX orientation at $E_{c.m.}=915.8$~MeV, 
 where the heaviest nucleus ($^{265}$Lr) is formed.
As described in section~\ref{subsec:numerical}, 
these non-central collisions are performed in a twice bigger box 
to avoid any spurious effect of the box before the full re-separation of the fragments.
Like in the central collision case, the nuclei start initially on the $x$ axis.
However, their initial Rutherford trajectory is determined for a finite impact parameter~$b$.

\begin{figure}[h]
\begin{center}
\subfloat[\label{Fig:ImpactParam-Proton}]{\includegraphics[width=0.42\textwidth]{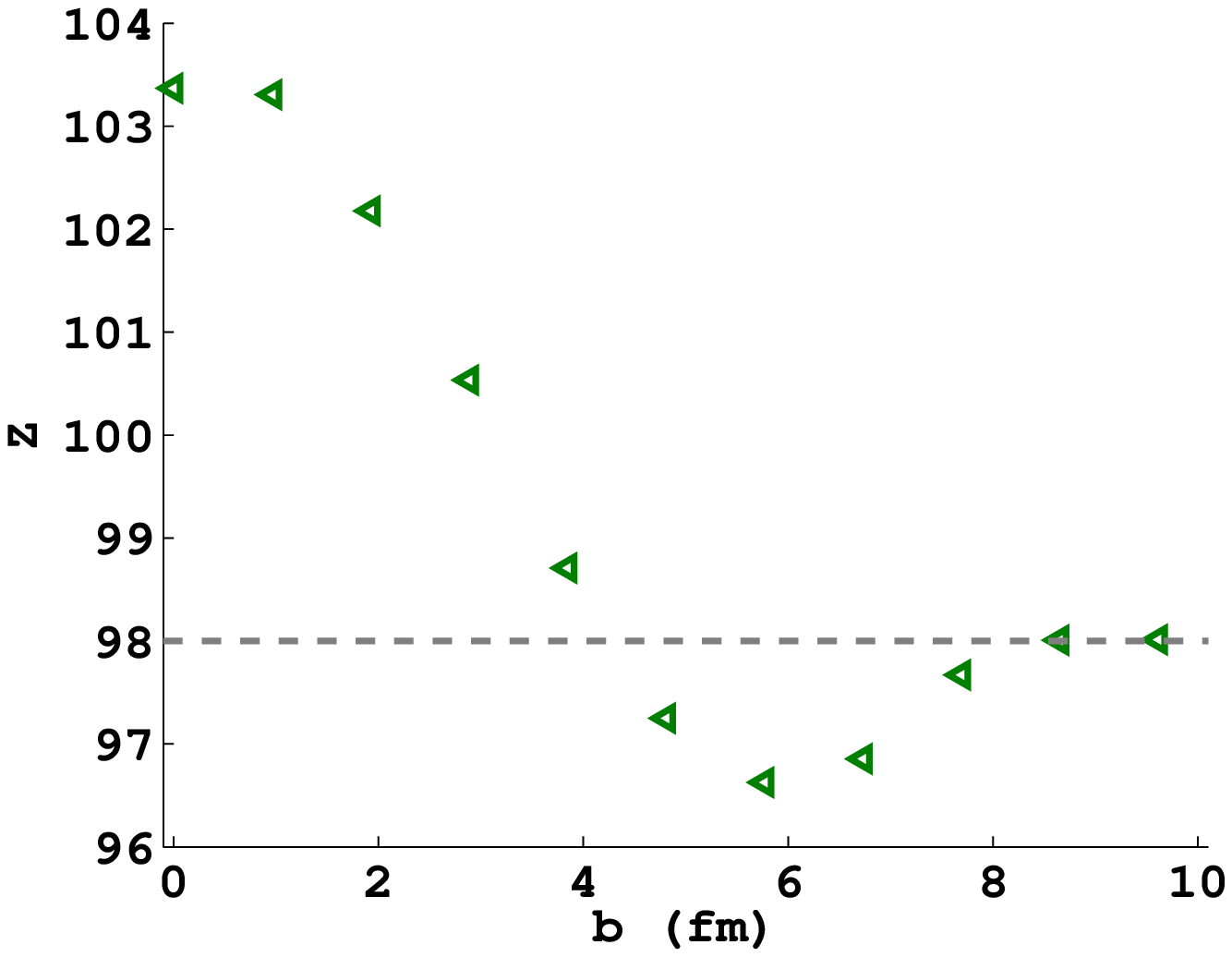}}\\
\subfloat[\label{Fig:ImpactParam-Neutron}]{\includegraphics[width=0.42\textwidth]{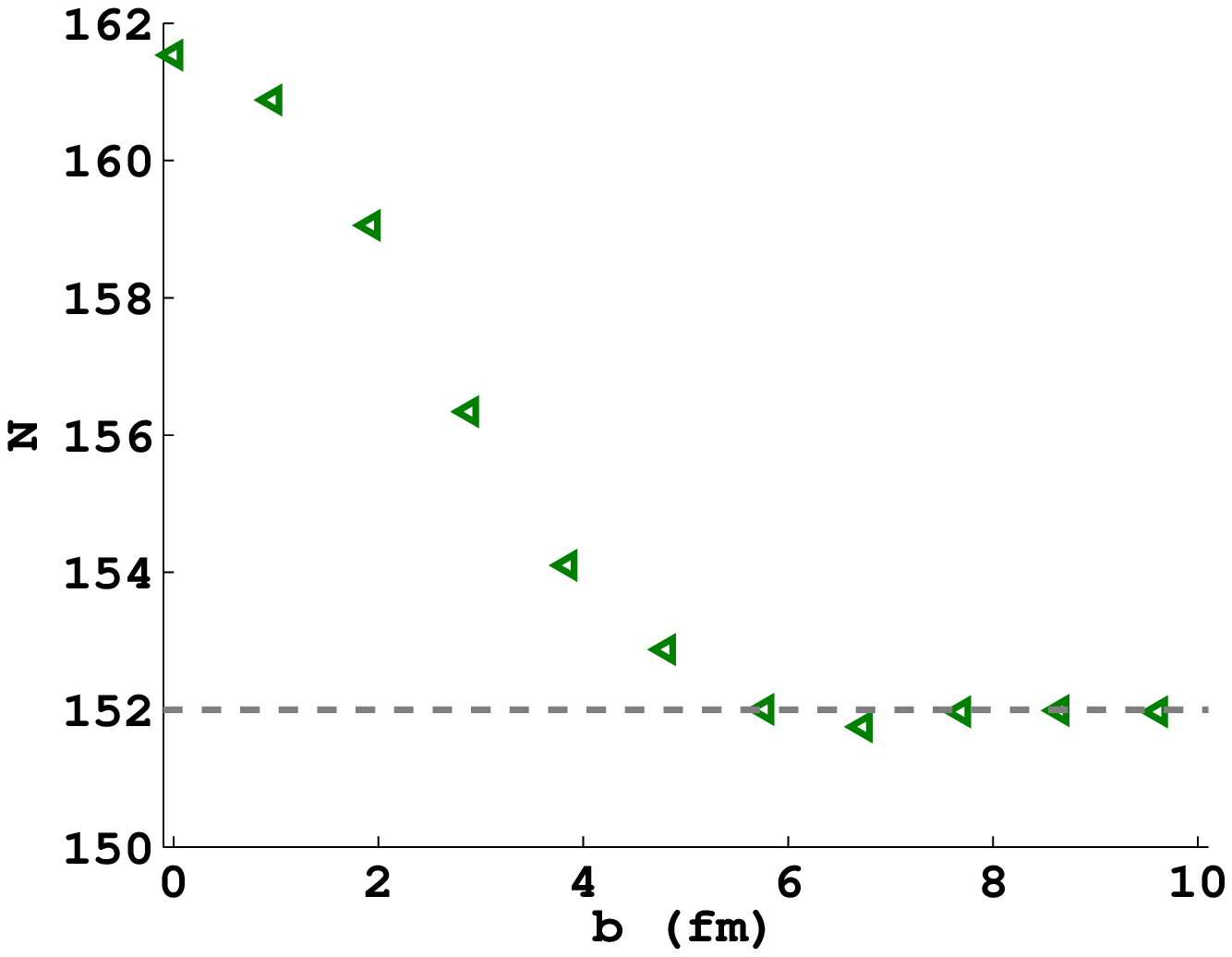}}\\
\end{center}
\caption{\label{Fig:ImpactParam}(color online). Number of  (a) protons $Z$ and (b) neutrons $N$ in the heavier fragment as function of the impact parameter at a center of mass energy $E_{c.m.}=915.8$~MeV in the YX relative orientation.
$^{250}$Cf is represented by the dashed lines at $\mathrm{Z}=98$ and $\mathrm{N}=152$.}
\end{figure}

Figs.~\ref{Fig:ImpactParam-Proton} and~\ref{Fig:ImpactParam-Neutron} 
display the post-collision number of protons and neutrons 
in the heavier fragment, respectively. 
The global effect of increasing the impact parameter is to reduce the number of nucleons transferred via IQ.
This can be interpreted in terms of a reduction of the contact time because of the centrifugal potential.
Note that the transfermium production is predicted to be dominant for this configuration up to $b\simeq3$~fm, 
corresponding to an angular momentum of $\sim118\hbar$.

It is interesting to note that IQ disappears for impact parameters above $b\simeq4$~fm.
In particular, the heavy fragments loses about one proton, but no neutron, for $4\le~b\le8$~fm. 
This can be understood in terms of charge equilibration 
as $^{250}$Cf is slightly more proton-rich than $^{232}$Th
and a transfer of one proton is enough to equilibrate this asymmetry. 
In addition, with the initial condition  for a YX orientation, 
a non-zero impact parameter shifts the system at contact, going away from the configuration
where the tip of one nucleus collides with the side of the other, which we identified
as the most favorable in terms of heavy elements production in section~\ref{subsec:multinucleon}. 
Finally, at $b>8$~fm, the overlap is not sufficient to allow any transfer of nucleons.

\section{Conclusions}
\label{sec:conclusion}

The time-dependent Hartree-Fock theory has been used to study 
the reaction mechanisms in the $^{232}$Th+$^{250}$Cf reaction.
The role of the deformation and relative orientation has been investigated 
and number of transfered nucleons, collision time, density in the overlap region and fast neutron emission have been analyzed.

A new process of inverse quasifission has been identified when the tip of the lighter nucleus collides with the side of the heavier one.
In this case, nucleons are transfered to the heavier nucleus and new neutron-rich transfermium nuclei can be produced. 
With the present reaction, $^{265}$Lr, which has three more neutrons than the most neutron-rich observed lawrencium isotope, could be produced in this process. In addition, fluctuations in the fragment neutron distribution should produce even more neutron-rich nuclei.

\begin{acknowledgments}
The authors are grateful to D.~J.~Hinde and M.~Dasgupta for stimulating discussions during this work and a careful reading of the paper.
D.~K. acknowledges the Australian National University summer school program during which this work has been done. 
The calculations have been performed on the Centre de Calcul Recherche et Technologie of the Commissariat \`a l'\'Energie Atomique, France, and on the NCI National Facility in Canberra,
Australia, which is supported by the Australian Commonwealth Government.
\end{acknowledgments}


\end{document}